\documentclass[aps,twocolumn,notitlepage,amsmath,amssymb,superscriptaddress]{revtex4-2}
\usepackage[utf8]{inputenc}
\usepackage{dcolumn}
\usepackage{amsthm}
\usepackage{bm}
\usepackage{graphicx,graphics,times,bm,bbm,amssymb,amsmath,amsthm,mathrsfs,MnSymbol}
\usepackage{gensymb}
\usepackage{amsfonts}
\usepackage{amsmath, amssymb}
\usepackage{times}
\usepackage{epsfig}
\usepackage{ulem} 
\usepackage[T1]{fontenc}
\pdfoutput=1
\usepackage[hypertexnames=false,pdfstartview=FitH]{hyperref}

\hypersetup{
    colorlinks=true,       
    linkcolor=magenta,          
    citecolor=blue,        
    filecolor=magenta,      
    urlcolor=blue,           
    runcolor=cyan}
\usepackage[pdftex]{color}
\usepackage{braket}
\newcommand{\po}[2]{\hat{\sigma}_{#1}^{#2}}
\usepackage{enumerate}
\usepackage[usenames,dvipsnames]{xcolor}
\usepackage{multirow}
\usepackage{mathtools}

\newcommand{\ignore}[1]{}

\usepackage{changes}
\newcommand{\beginsupplement}{%
  \renewcommand\figurename{Supplementary Figure}
  \renewcommand\tablename{Supplementary Table}
  \setcounter{table}{0}
  \renewcommand{\thetable}{~\arabic{table}}%
  \setcounter{figure}{0}
  \renewcommand{\thefigure}{~\arabic{figure}}%
  \setcounter{section}{0}
  \setcounter{equation}{0}
  \setcounter{page}{1}
}

\begin{document}

\title{Non-Equilibrium Criticality-Enhanced Quantum Sensing with Superconducting Qubits}

\author{Hao Li}
\altaffiliation[]{These authors contributed equally to this work.}
\affiliation{Beijing Key Laboratory of Fault-Tolerant Quantum Computing, Beijing Academy of Quantum Information Sciences, Beijing 100193, China}

\author{Yaoling Yang}
\altaffiliation[]{These authors contributed equally to this work.}
\affiliation{Institute of Fundamental and Frontier Sciences, University of Electronic Science and Technology of China, Chengdu 611731, China}

\author{Yun-Hao Shi}
\affiliation{Beijing National Laboratory for Condensed Matter Physics, Institute of Physics, Chinese Academy of Sciences, Beijing 100190, China}

\author{Zheng-An Wang}
\affiliation{Beijing Key Laboratory of Fault-Tolerant Quantum Computing, Beijing Academy of Quantum Information Sciences, Beijing 100193, China}

\author{Ziting Wang}
\affiliation{Beijing Key Laboratory of Fault-Tolerant Quantum Computing, Beijing Academy of Quantum Information Sciences, Beijing 100193, China}

\author{Jintao Li}
\affiliation{Beijing Key Laboratory of Fault-Tolerant Quantum Computing, Beijing Academy of Quantum Information Sciences, Beijing 100193, China}

\author{Yipeng Zhang}
\affiliation{Beijing Key Laboratory of Fault-Tolerant Quantum Computing, Beijing Academy of Quantum Information Sciences, Beijing 100193, China}

\author{Kui Zhao}
\affiliation{Beijing Key Laboratory of Fault-Tolerant Quantum Computing, Beijing Academy of Quantum Information Sciences, Beijing 100193, China}

\author{Yue-Shan Xu}
\affiliation{Beijing Key Laboratory of Fault-Tolerant Quantum Computing, Beijing Academy of Quantum Information Sciences, Beijing 100193, China}

\author{Cheng-Lin Deng}
\affiliation{Beijing Key Laboratory of Fault-Tolerant Quantum Computing, Beijing Academy of Quantum Information Sciences, Beijing 100193, China}

\author{Yu Liu}
\affiliation{Beijing National Laboratory for Condensed Matter Physics, Institute of Physics, Chinese Academy of Sciences, Beijing 100190, China}

\author{Wei-Guo Ma}
\affiliation{Beijing National Laboratory for Condensed Matter Physics, Institute of Physics, Chinese Academy of Sciences, Beijing 100190, China}
\affiliation{School of Physical Sciences, University of Chinese Academy of Sciences, Beijing 100049, China}

\author{Tian-Ming Li}
\affiliation{Beijing National Laboratory for Condensed Matter Physics, Institute of Physics, Chinese Academy of Sciences, Beijing 100190, China}
\affiliation{School of Physical Sciences, University of Chinese Academy of Sciences, Beijing 100049, China}

\author{Jiachi Zhang}
\affiliation{Beijing National Laboratory for Condensed Matter Physics, Institute of Physics, Chinese Academy of Sciences, Beijing 100190, China}
\affiliation{School of Physical Sciences, University of Chinese Academy of Sciences, Beijing 100049, China}

\author{Cai-Ping Fang}
\affiliation{Beijing National Laboratory for Condensed Matter Physics, Institute of Physics, Chinese Academy of Sciences, Beijing 100190, China}
\affiliation{School of Physical Sciences, University of Chinese Academy of Sciences, Beijing 100049, China}

\author{Jia-Cheng Song}
\affiliation{Beijing National Laboratory for Condensed Matter Physics, Institute of Physics, Chinese Academy of Sciences, Beijing 100190, China}
\affiliation{School of Physical Sciences, University of Chinese Academy of Sciences, Beijing 100049, China}

\author{Hao-Tian Liu}
\affiliation{Beijing National Laboratory for Condensed Matter Physics, Institute of Physics, Chinese Academy of Sciences, Beijing 100190, China}
\affiliation{School of Physical Sciences, University of Chinese Academy of Sciences, Beijing 100049, China}

\author{Si-Yun Zhou}
\affiliation{Beijing National Laboratory for Condensed Matter Physics, Institute of Physics, Chinese Academy of Sciences, Beijing 100190, China}
\affiliation{School of Physical Sciences, University of Chinese Academy of Sciences, Beijing 100049, China}

\author{Zheng-He Liu}
\affiliation{Beijing National Laboratory for Condensed Matter Physics, Institute of Physics, Chinese Academy of Sciences, Beijing 100190, China}
\affiliation{School of Physical Sciences, University of Chinese Academy of Sciences, Beijing 100049, China}

\author{Bing-Jie Chen}	
\affiliation{Beijing National Laboratory for Condensed Matter Physics, Institute of Physics, Chinese Academy of Sciences, Beijing 100190, China}
\affiliation{School of Physical Sciences, University of Chinese Academy of Sciences, Beijing 100049, China}

\author{Gui-Han Liang}
\affiliation{Beijing National Laboratory for Condensed Matter Physics, Institute of Physics, Chinese Academy of Sciences, Beijing 100190, China}

\author{Xiaohui Song}
\affiliation{Beijing National Laboratory for Condensed Matter Physics, Institute of Physics, Chinese Academy of Sciences, Beijing 100190, China}

\author{Zhongcheng Xiang}
\email{zcxiang@iphy.ac.cn}
\affiliation{Beijing National Laboratory for Condensed Matter Physics, Institute of Physics, Chinese Academy of Sciences, Beijing 100190, China}
\affiliation{School of Physical Sciences, University of Chinese Academy of Sciences, Beijing 100049, China}
\affiliation{Hefei National Laboratory, Hefei 230088, China}

\author{Kai Xu}
\affiliation{Beijing National Laboratory for Condensed Matter Physics, Institute of Physics, Chinese Academy of Sciences, Beijing 100190, China}
\affiliation{School of Physical Sciences, University of Chinese Academy of Sciences, Beijing 100049, China}
\affiliation{Beijing Key Laboratory of Fault-Tolerant Quantum Computing, Beijing Academy of Quantum Information Sciences, Beijing 100193, China}
\affiliation{Hefei National Laboratory, Hefei 230088, China}
\affiliation{Songshan Lake Materials Laboratory, Dongguan 523808, Guangdong, China}

\author{Kaixuan Huang}
\email{huangkx@baqis.ac.cn}
\affiliation{Beijing Key Laboratory of Fault-Tolerant Quantum Computing, Beijing Academy of Quantum Information Sciences, Beijing 100193, China}

\author{Abolfazl Bayat}
\email{abolfazl.bayat@uestc.edu.cn}
\affiliation{Institute of Fundamental and Frontier Sciences, University of Electronic Science and Technology of China, Chengdu 611731, China}
\affiliation{Key Laboratory of Quantum Physics and Photonic Quantum Information, Ministry of Education, University of Electronic
Science and Technology of China, Chengdu 611731, China}
\affiliation{Shimmer Center, Tianfu Jiangxi Laboratory, Chengdu 641419, China}

\author{Heng Fan}
\email{hfan@iphy.ac.cn}
\affiliation{Beijing National Laboratory for Condensed Matter Physics, Institute of Physics, Chinese Academy of Sciences, Beijing 100190, China}
\affiliation{School of Physical Sciences, University of Chinese Academy of Sciences, Beijing 100049, China}
\affiliation{Beijing Key Laboratory of Fault-Tolerant Quantum Computing, Beijing Academy of Quantum Information Sciences, Beijing 100193, China}
\affiliation{Hefei National Laboratory, Hefei 230088, China}
\affiliation{Songshan Lake Materials Laboratory, Dongguan 523808, Guangdong, China}

\begin{abstract}
\textbf{ Exploiting quantum features allows for estimating external parameters with precisions well beyond the capacity of classical sensors, a phenomenon known as quantum-enhanced precision. Quantum criticality has been identified as a resource for achieving such enhancements with respect to the probe size. However, they demand complex probe preparation and measurement and the achievable enhancement is ultimately restricted to narrow parameter regimes. On the other hand, non-equilibrium probes harness dynamics, enabling  quantum-enhanced precision with respect to time over a wide range of parameters through simple probe initialization. Here, we unify these approaches through a Stark-Wannier localization platform, where competition between a linear gradient field and particle tunneling enables quantum-enhanced sensitivity across an extended parameter regime. The probe is implemented on a 9-qubit superconducting quantum device, in both single- and double-excitation subspaces, where we explore its performance in the extended phase, the critical point and the localized phase. Despite employing only computational-basis measurements we have been able to achieve near-Heisenberg-limited precision by combining outcomes at distinct evolution times.  In addition, we demonstrate that the performance of the probe in the entire extended phase is significantly outperforming the performance in the localized regime. Our results highlight Stark-Wannier systems as versatile platforms for quantum sensing, where the combination of criticality and non-equilibrium dynamics enhances precision over a wide range of parameters without stringent measurement requirements. }
\end{abstract}

\maketitle

\noindent
\textbf{\large{INTRODUCTION}}\\
\noindent
Quantum sensors can outperform their classical counterparts by harnessing quantum features, such as superposition and entanglement~\cite{Giovannetti2004Quantum-Enhanced,Giovannetti2011Advances,Degan2017quantum,Ye2024Essay,agarwal2025quantumsensing}. Their superiority is quantified by sensing precision, measured through the estimation variance, which scales as $N^{-\beta}$, where $N$ represents resources like probe size $L$ or time $t$. While in the absence of quantum features the precision is limited to shot noise ($\beta{=}1$), leveraging quantum features allows for achieving $\beta{>}1$, known as quantum-enhanced precision~\cite{paris2009quantum}. Many-body systems, in or out of equilibrium, are promising platforms for such enhancement~\cite{Montenegro2025Review}. Only recently, experimental demonstrations of such quantum probes have been reported in Rydberg atoms~\cite{Ding2022enhanced}, solid state systems~\cite{liu2021experimental,Wu2024experimental,moon2024discrete}, photonic setups~\cite{xiao2025observation} and superconducting devices~\cite{yu2025experimental,Beaulieu2025criticalityEnhanced}. Equilibrium probes exploit different forms of quantum criticality near phase transitions~\cite{CamposVenuti2007quantum,Zanardi2008quantum,Invernizzi2008optimal,Rams2018AtTheLimit,Chu2021dynamic,sarkar2022freefermionic,Sarkar2025Exponentially}, achieving $L^{-\beta}$ (with $\beta{>}1$) scaling but only in a narrow parameter regime with complex probe initialization and demanding sophisticated measurement setups~\cite{Garbe2020critical,Gietka2021adiabatic,Abiuso2025fundamental}. In contrast, non-equilibrium probes utilize the dynamics and harness evolution time $t$ as a resource and benefit from simpler initialization. They achieve ${\sim}1/t^\beta$ scaling for their estimation variance where $\beta{>}1$, i.e.\ quantum-enhanced sensitivity, is not limited to a narrow region and can even reach $\beta{=}2$ (Heisenberg limit), when the probe is a closed system under unitary evolution and measurement is optimal~\cite{boixo2007generalized,Pang2014quantum}. Several open questions arise: (i) Is it possible to combine criticality-enhanced quantum sensing with non-equilibrium dynamics to benefit from both resources, namely probe size and time, over a wide range of parameters? (ii) Is it feasible to demonstrate quantum-enhanced sensitivity with respect to time in near-term quantum devices which are prone to decoherence and only simple measurements are available?  

Unlike conventional quantum phase transitions (e.g., first- and second-order), which only affect the ground state, localization transitions occur across the entire spectrum. Stark-Wannier localized systems~\cite{Wannier1960wave}, where a linear gradient field competes with tunneling in the Hamiltonian, exemplify this behavior, with non-equilibrium signatures like Bloch oscillations~\cite{bloch1929quantenmechanik, Guo_bloch_2021,Guo2021stark,Ferrari2006LongLived}. Recently, Stark-Wannier systems have emerged as promising candidates for quantum-enhanced sensing of gradient fields~\cite{He2023stark,manshouri2024quantum}. 
In this work, we experimentally realize a non-equilibrium Stark-Wannier quantum sensor using a 9-qubit superconducting quantum circuit to estimate gradient field strengths. We explore the sensing capacity of the probe, in both single- and double-excitation subspaces, across the extended phase, the critical point, and the localized phase. While achieving quantum-enhanced precision typically requires an optimal measurement basis that depends on both time and the gradient field, our experiments employ a simplified approach by performing measurements only in the computational basis. Despite this, we demonstrate that combining measurement outcomes at different times enables near-Heisenberg-limited precision without the need for complex measurement basis. Our results show that the performance of the probe is significantly better in the extended phase in comparison with the localized phase, evidencing the criticality-enhanced sensing. \\

\begin{figure*}[t]
	\centering
	\includegraphics[width=0.8\linewidth]{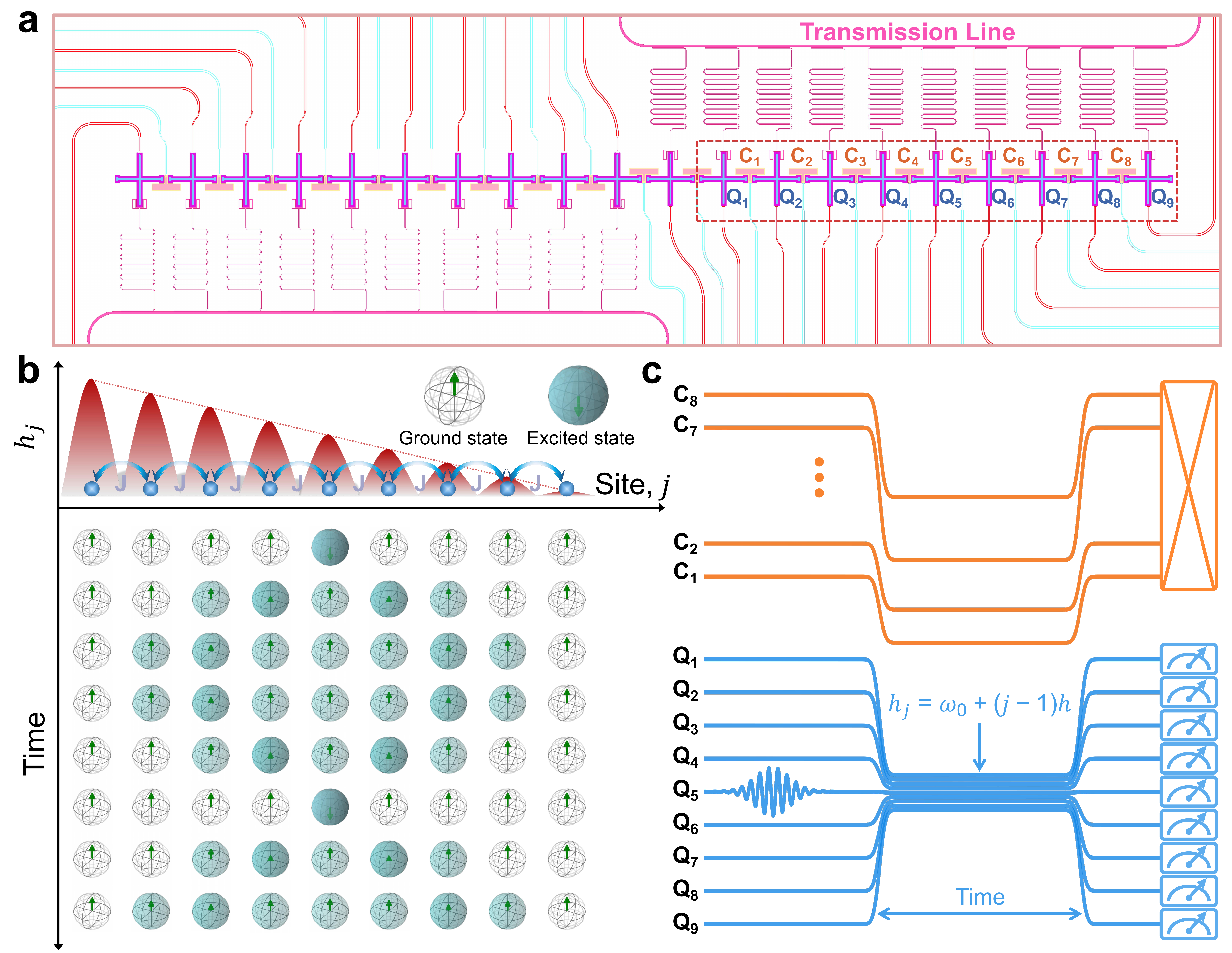}\\
	\caption{\textbf{Quantum device and experimental scheme.} \textbf{a}, False-color optical micrograph of the 20-qubit superconducting quantum device, with the dashed rectangle indicating 9 transmon qubits Q$_1$-Q$_9$ and 8 couplers C$_1$-C$_8$ employed experimentally. \textbf{b}, Particle transport under Stark-Wannier gradient field. The upper panel is a schematic representation of the Stark-Wannier gradient field with both the on-site and off-diagonal quasi-periodic modulations. The nearest-neighbor coupling strength $J$ was set to $-8.0\,\textrm{MHz}$, and the gradient field $h$ was tuned from $0\,\textrm{MHz}$ to $-30.0\,\textrm{MHz}$. The lower panel shows the excitation propagation from the central site over time when the gradient field strength is near the phase transition point. In this situation, Bloch oscillations, which occur in the localized phase, cause the excitation to delocalize across the entire system and then bring it back to its initial position.     
    Dark gray spheres represent excited states, and transparent spheres represent ground states.  \textbf{c}, Experimental pulse sequences for observing quantum-enhanced sensitivity in the Stark-Wannier system with single excitation. The central qubit Q$_5$ was excited to the $\ket{1}$ state using an $X_\pi$ pulse, followed by precisely tuned evolution under gradient $h_j$ and coupling $J$ for duration $t$, with final measurement of the population probabilities of the $|1\rangle$ state across all qubits.}
    \label{Fig1_schematic}
\end{figure*}

\noindent
\textbf{\large{RESULTS}}\\
\noindent
\textbf{Estimation theory} \\
\noindent
Quantum sensing deals with extracting information about an unknown parameter $h$ from a  probe which is described by quantum state $\rho_h$. To achieve this, one has to perform a measurement described by positive operator-valued measure (POVM) projectors $\{\Pi_n\}$. Each outcome is observed with probability $p_n(h) {=} \text{Tr}[\rho_h \Pi_n]$. By post-processing the measurement outcomes one can construct an unbiased estimator $\hat{h}$ whose average is taken as the estimation $h_{\rm est}{=}\langle \hat{h} \rangle $ for the real value $h$. The precision of the estimation by using $\mathcal{M}$ measurement samples, characterized by the variance $\delta^2 h$ of the estimator $\hat{h}$, is constrained by the  Cram\'er-Rao inequality $\delta^2 h {\geq} 1/\mathcal{M} \mathcal{F}_C {\geq} 1/\mathcal{M} \mathcal{F}_Q$~\cite{paris2009quantum}. Here, the classical Fisher information (CFI) $\mathcal{F}_C = \sum_n (\partial_h p_n)^2 / p_n$ sets the achievable precision for a specific POVM $\{\Pi_n\}$, while the quantum Fisher information (QFI) $\mathcal{F}_Q$ represents the ultimate achievable precision which is obtained by maximizing the CFI with respect to all possible measurement outcomes. For pure states $\rho_h {=} \ket{\psi_h} \bra{\psi_h}$, the QFI simplifies to $\mathcal{F}_Q {=} 4\left(\braket{\partial_h \psi_h|\partial_h \psi_h} - |\braket{\partial_h \psi_h|\psi_h}|^2 \right)$~\cite{paris2009quantum}.\\

\noindent
\textbf{Model and set-up} \\
\noindent
Superconducting quantum systems are among the most promising platforms for demonstrating practical quantum computing capabilities due to their scalability, precise control over individual qubits and their interactions, and high-fidelity measurements~\cite{Roushan2017, gong_quantum_2021, Guo_MBL_2020, Ni_EVP_2023, Cai2024, Storz2023, Shtanko2025, Robledo2025, Andersen2025, Google_EC_2025, Zuchongzi3_2025, alghadeer2025lowcrosstalkscalablesuperconducting}. Our quantum device features a one-dimensional architecture with $20$ frequency-tunable transmon qubits arranged in a linear configuration, interconnected through $19$ couplers mediating inter-qubit interactions~\cite{yan_tunable_2018}, see Fig.\ref{Fig1_schematic}\textbf{a} and check the Supplementary Materials (SM) Note 1 for the fabrication details. We use a chain array of $L{=}9$ adjacent qubits in the experiments. The remaining qubits and couplers are spectrally detuned below $3.5\,\textrm{GHz}$ through flux-biased tuning, maintaining at least $800\,\textrm{MHz}$ separation from the idle frequencies of the experimentally activated qubits ($4.3{-}5.0\,\textrm{GHz}$) to avoid unwanted interactions. 
By setting $\hbar {=} 1$, the effective Hamiltonian of the  system reads~\cite{xiang_simulating_2023}:
\begin{equation}
   \frac{\hat {H}}{2\pi} =J \sum\limits_{ j =1 } ^{L-1}\left(\po{j}{+}\po{j+1}{-}+\po{j}{-}\po{j+1}{+}\right)+ \sum\limits_{ j =1 } ^{L} h_j\po{j}{+}\po{j}{-},
  \label{eq_Ham}
\end{equation}
where $\po{j}{+}$ ($\po{j}{-}$) denotes the raising (lowering) operator, $J{=}{-}8$ MHz is the exchange coupling, $h_j{=}(j{-}1)h$ is the Stark linear potential of the qubit and $h$ is the gradient field to be estimated. The qubit's on-site potential is modulated as $\omega_j/2\pi{=}\omega_0/2\pi {+} h_j$, with $\omega_0/2\pi {=} 4.5$ \textrm{GHz} being the reference frequency. The relevant schematic diagram is shown in the upper panel of Fig.~\ref{Fig1_schematic}\textbf{b}. In this Hamiltonian, varying the $h{/}J$ ratio drives the system through a phase transition from an extended phase to a Stark-Wannier localized phase~\cite{Wannier1960wave}. The details about the calibration procedures can be found in the SM, see Note 2. As $h{/}J$ increases, the system goes from an extended phase, where all the eigenstates are thermalized, to a localized phase, where all the eigenstates are localized. Since this localization transition occurs throughout the entire spectrum, its features can be observed in non-equilibrium dynamics through Bloch oscillations~\cite{bloch1929quantenmechanik, Guo_bloch_2021,Guo2021stark}. The schematic diagram of the evolution near the phase transition point is shown in the lower panel of Fig.~\ref{Fig1_schematic}\textbf{b}. Such dynamics can also be used for inferring the gradient field $h$ with quantum-enhanced sensitivity~\cite{manshouri2024quantum}. Here, we experimentally demonstrate such enhanced quantum precision for sensing the gradient field $h$ relying on simple experimentally available measurements. \\

We first consider a probe with a single excitation described by the basis $\ket{\mathbf{j}}{=}\ket{0,\cdots0,1,0,\cdots,0}$ in which all qubits are in the state $\ket{0}$ except qubit $j$ which is in the state $\ket{1}$. The system is initialized by creating single excitation at the middle of the chain, namely $\ket{\Psi(0)}{=}\ket{\mathbf{5}}$ (note that system size is $L{=}9$). The system is evolved under the action of the Hamiltonian~(\ref{eq_Ham}) as
\begin{equation}
    \ket{\Psi_h(t)}=e^{-i\hat{H}t}\ket{\Psi(0)}=\sum_{j=1}^L c_j(t,h) \ket{\mathbf{j}}
    \label{eq_time_evolution}
\end{equation}
where, $c_j(t,h){=}\langle \mathbf{j}|e^{-i\hat{H}t}|\Psi(0)\rangle$ encodes information about the parameter of interest, i.e. $h$. The evolution is experimentally implemented through following the waveform sequence depicted in Fig.~\ref{Fig1_schematic}\textbf{c}. 
Then we measure the site-resolved on-site population of each qubit $P_j(t,h){=}|c_j(t,h)|^2$ across a $0{-}350\,\textrm{ns}$ evolution window, executing $10$ experimental
repetitions, with each repetition covering $5000$ single-shot quantum non-demolition readouts. By increasing $h$ the system makes a transition from an extended phase to a localized phase. In the SM, see Note 3, we discuss how to identify the transition point which separates the extended phase from the localized phase.\\

\begin{figure*}[t]
	\centering
	\includegraphics[width=0.8\linewidth]{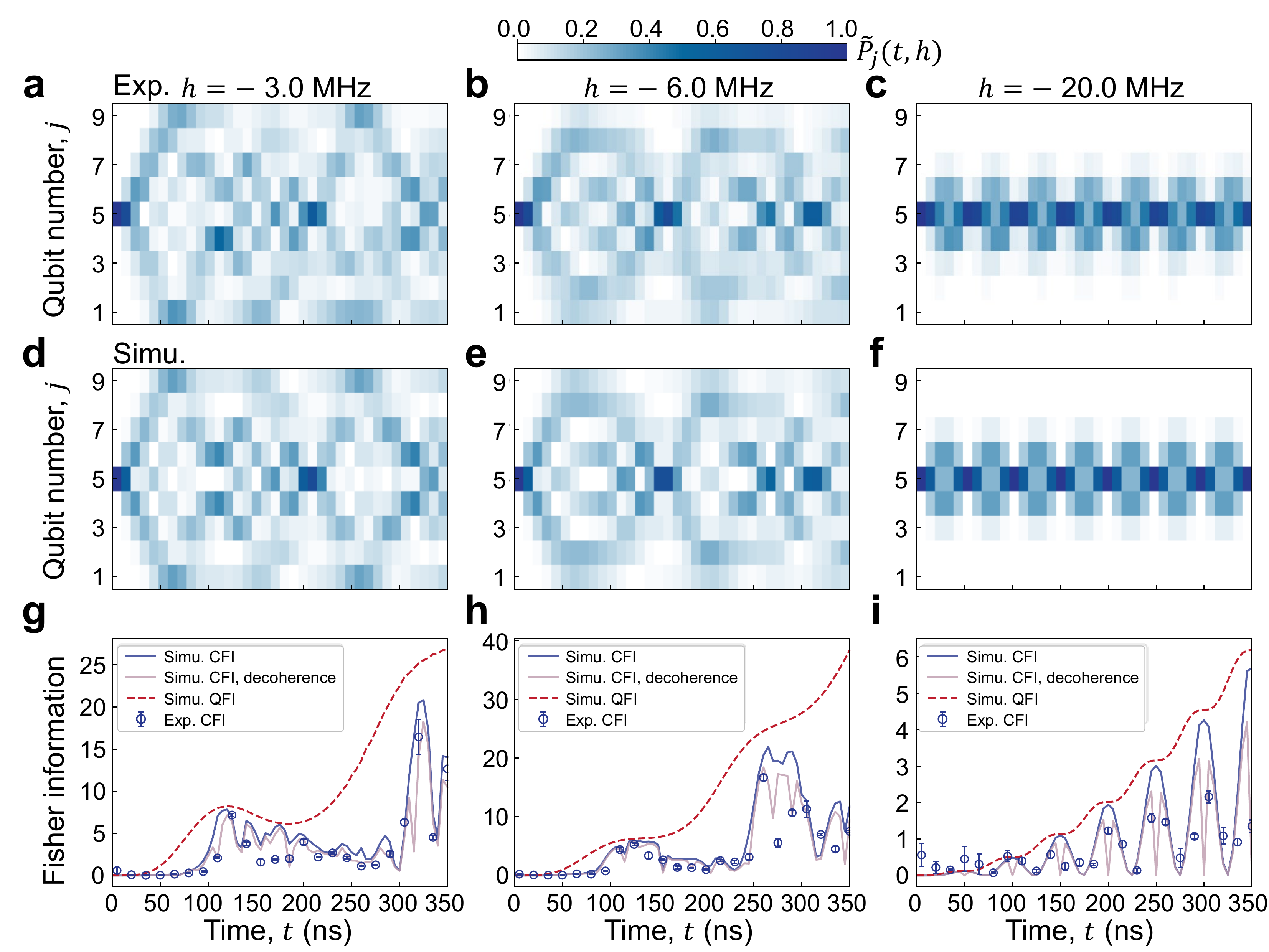}\\
	\caption{\textbf{Excitation transport and Fisher information.} The time evolution of the empirical on-site population $\tilde{P}_j$, statistically averaged over $10$ independent repetitions with each repetition covering $5000$ single shots, with the system initialized in $|\Psi(0)\rangle {=} |\mathbf{5}\rangle$ for: \textbf{a}, the extended phase with $h {=} {-}3.0\,\textrm{MHz}$; \textbf{b}, around the transition point with $h {=} {-}6.0\,\textrm{MHz}$; and \textbf{c}, the localized phase with $h {=} {-}20.0\,\textrm{MHz}$. The corresponding numerical simulations for the experimental data shown in panels \textbf{a-c} are depicted in panels \textbf{d-f}.  \textbf{g-i}, Fisher information corresponding to $h {=} {-}3.0\,\textrm{MHz}$, $h {=} {-}6.0\,\textrm{MHz}$ and $h {=} {-}20.0\,\textrm{MHz}$ respectively, with error bars representing the one standard deviation (1 SD) derived from $25$ experimental repetitions with each repetition covering $10000$ single shots. Fisher information is expressed in units of $\mathrm{MHz}^{-2}$.}
    \label{Fig2_transport_and_FI}
\end{figure*}

\noindent
\textbf{Stark-Wannier field sensing} \\
\noindent
In order to show the performance of our device, in Fig.~\ref{Fig2_transport_and_FI}\textbf{a-c} we plot the experimentally measured population distribution $P_j(t,h)$ as a function of time $t$ and site index $j$ for $h{=} {-}3$ MHz (extended phase), $h{=} {-}6$ MHz (around the transition point) and $h{=}{-}20$ MHz (localized phase), respectively. Their corresponding numerical simulations are also plotted in  Fig.~\ref{Fig2_transport_and_FI}\textbf{d-f} which show a very good match between experiment and simulation. In the extended regime, see Fig.~\ref{Fig2_transport_and_FI}\textbf{a}, the excitation is dispersed across the whole system and only partially returns to its original site after reflection from the boundaries. On the other hand, around the transition point, Bloch oscillation is observed in the system in which the excitation returns to its original site almost perfectly after delocalizing in the entire system, see Fig.~\ref{Fig2_transport_and_FI}\textbf{b}. By increasing $h$, the system undergoes a transition to the localized phase, in which the Bloch oscillations occur very locally among a few sites at the middle of the chain, see Fig.~\ref{Fig2_transport_and_FI}\textbf{c}.  \\

In our experiment, we rely on measuring the qubits in the computational basis, which is described by the projectors $\{\Pi_{j}{=}\ket{\mathbf{j}}\bra{\mathbf{j}} \}$. The goal is to use these measurement outcomes and estimate the gradient field $h$. Since the measurement basis is fixed, according to the Cram\'er-Rao inequality, the precision is bounded by the CFI. In order to obtain the CFI in our experiment, we have to first get the empirical probabilities $\tilde{P}_j(t,h){=}n_j/n$, where $n_j$ is the number of times that the outcome $\ket{\mathbf{j}}$ is obtained and $n{=}\sum_{j=1}^L n_j$ is the total number of samplings. In the large  $n$ limit, the empirical probabilities converge to the actual probabilities, namely $\lim_{n\to \infty}\tilde{P}_j(t,h){=}P_j(t,h)$. Therefore, the empirical evaluation of the  CFI will be
\begin{equation}
   \mathcal{F}_C(t,h) = \sum_{j=1}^L \left[\left(\frac{\tilde{P}_j(t,h)-\tilde{P}_j(t, h-\varepsilon)}{\varepsilon}\right)^2 \frac{1}{\tilde{P}_j(t, h)}\right].
  \label{eq_CFI}
\end{equation}
Here, we set $\varepsilon {=} 0.1\,\textrm{MHz}$, which ensures the balance between precision requirements for gradient sensitivity quantification and the resolvable field-interval threshold of the transport experiment. Experimental results originate from $25$ independent datasets, with each dataset comprising $10000$ single-shot measurements, resulting in a total sample size of $n{=}2.5{\times}10^5$. In Fig.~\ref{Fig2_transport_and_FI}\textbf{g-i}, we plot the temporal evolution of the CFI estimated from the empirical data as well as the theoretical simulations for the three corresponding regimes, namely extended ($h{=}{-}3$ MHz), near the transition ($h{=}{-}6$ MHz) and localized ($h{=}{-20}$ MHz). In the same figures, we also plot the QFI, obtained through numerical simulation, which is clearly the upper bound of the CFI obtained from both theory and experiment.  The CFI does not saturate the QFI because our simple measurement scheme is not always the optimal measurement basis. In addition, while the empirical and theoretical values of the CFI match well at short time scales, they start to deviate as time increases. This is because our evolution is ultimately an open quantum system dynamics where decoherence starts to take effect. In order to show this, in Fig.~\ref{Fig2_transport_and_FI}\textbf{g-i} we also plot the CFI obtained from numerical simulation of an open quantum dynamics in which dissipation and dephasing have been incorporated. As the figure clearly shows, by incorporating the decoherence effects the matching between the experimental data and theoretical predictions significantly improves. For the details of the open quantum system dynamics, please see the Methods.\\

\begin{figure*}[t]
	\centering
	\includegraphics[width=1\linewidth]{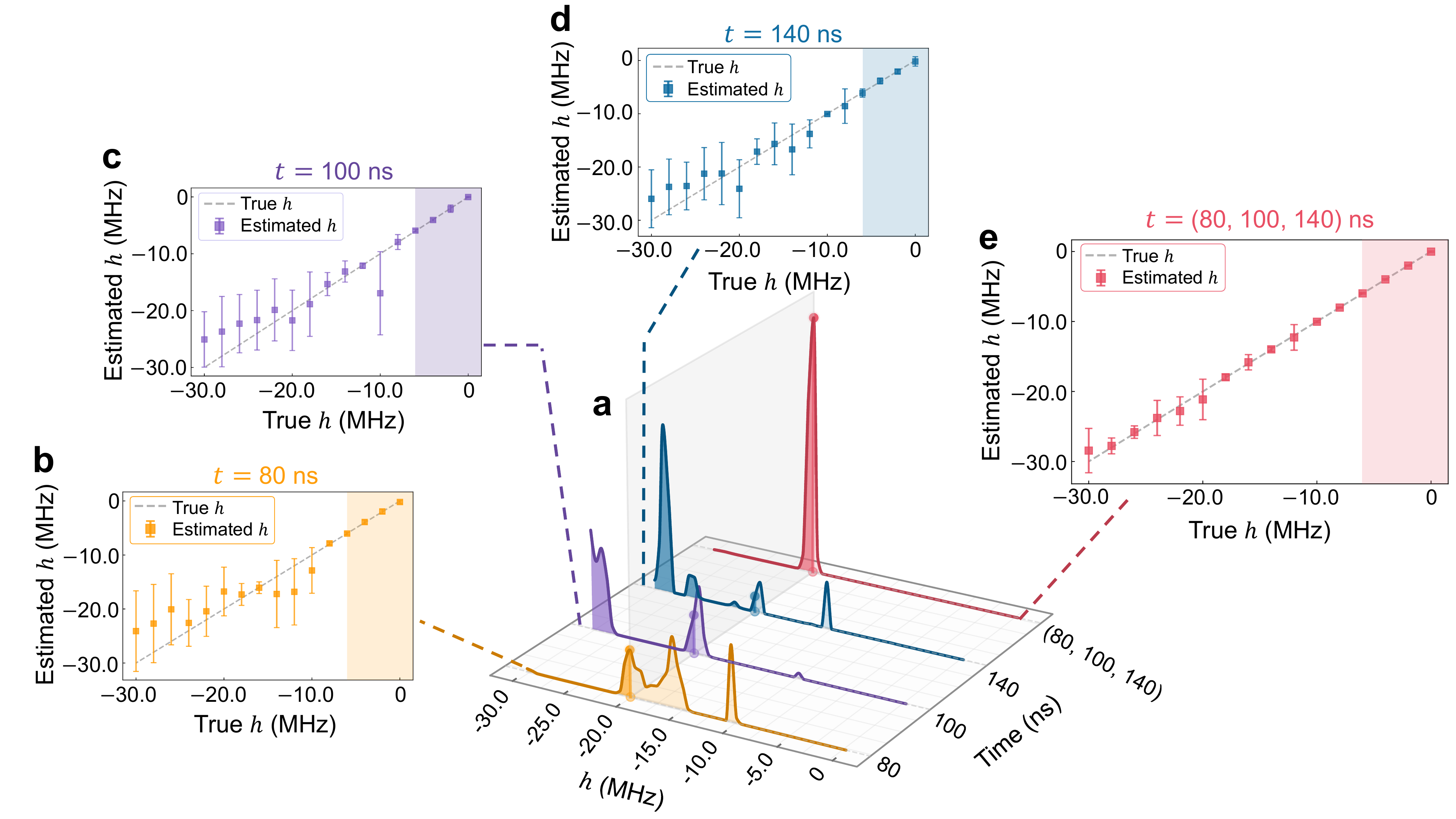}\\
	\caption{ \textbf{Bayesian estimation using single-time and three-time combined approaches.} \textbf{a}, Single-time posterior probability distributions at $t {=} 80$ (yellow line), $100$ (purple line), $140$~ns (blue line) using $\mathcal{M}{=}60$  samples and three-time combined (from the data obtained at $t{=}80$~ns, $t{=}100$~ns and $t{=}$140~ns) posterior distribution (red line) using the total sample size of $\mathcal{M}{=}60$ ($20$ samples from each time), respectively. The true value of the parameter, namely $h{=}{-}20$ MHz, is indicated by a gray plane.   
    The single-time results demonstrate multi-peak behavior in the posterior that leads to inaccurate parameter estimation, while the three-time combined posterior shows a single sharp peak that accurately identifies the true parameter value. \textbf{b-d}, Correspondence between the estimated value $h_{\rm est}$ and the true $h$, using single-time posterior,  within the range of $0$~MHz to $-30$~MHz, with error bars representing 1 SD. The dashed line indicates perfect estimation and the shaded regions denote the extended phase, where the error bars are smaller than those in the localized phase. \textbf{e}, The estimated value $h_{\rm est}$ versus true $h$, using the three-time combined posterior, which demonstrates significantly improved parameter estimation precision across the entire range of $h$.}
    \label{Fig3_BE_1excitation}
\end{figure*}

\noindent
\textbf{Bayesian estimation} \\
\noindent
While Fisher information indicates the bound for the achievable precision, in practice one has to adopt an estimator to accomplish a real estimation task. Here, we use a Bayesian estimator to infer the unknown parameter from total repeated measurement samples $\mathcal{M}$. Each measurement outcome appears $n_j$ times such that $\sum_jn_j{=}\mathcal{M}$. The Bayesian estimator calculates the posterior probability based on Bayes' rule, namely:
\begin{equation}
f(h|\{ n_j\}) = \frac{f(\{ n_j\}|h)f(h)}{f(\{ n_j\})},
\end{equation}
where $f(h|\{ n_j\})$ denotes the posterior probability distribution, $f(\{ n_j\}|h)$ represents the likelihood of the observed data given $h$, $f(h)$ is the prior information about $h$, and $f(\{ n_j\})$ is the normalization constant that ensures the posterior sums to 1. The estimated value is then determined by $h_{\rm est} = \arg\max_h f(h|\{ n_j\})$. See the Methods  for more details about our Bayesian estimation. 
Ideally, the posterior distribution should exhibit a Gaussian shape, which indicates the effectiveness of the estimation. In Fig.~\ref{Fig3_BE_1excitation}\textbf{a}, we plot the posterior probability distribution  for the sample size of $\mathcal{M}{=} 60$ when the true value of the field is $h{=}{-}20.0\,\textrm{MHz}$ at different times of $t {=} 80$ ns, $t{=}100$ ns, and $t{=}140$ ns, respectively. Contrary to expectations, the posterior distribution exhibits multi-peak behavior, indicating that multiple distinct values of $h$ can generate nearly identical measurement outcomes $\{n_j \}$. Thus, by relying on the measurement outcomes obtained at a single given time $t$, one cannot have a reliable estimation.  To address this, we partition the measurements into three groups, each repeated $\mathcal{M}/3$ times, ensuring the total measurement count $\mathcal{M}$ remains unchanged across the three specified time points. Because measurement outcomes at different times are statistically independent, the final posterior distribution is obtained by multiplying the individual posteriors at each specified time. The resulting posterior is also plotted in Fig.~\ref{Fig3_BE_1excitation}\textbf{a}, which shows a clear single peak posterior with a very reliable estimation. In Fig.~\ref{Fig3_BE_1excitation}\textbf{b-d}, we plot the estimated value  $h_{\rm est}$ as a function of the real value $h$, based on measurements at a single time. The error bars are obtained by repeating the procedure for $50$ times. As the figures clearly show, relying on single-time measurements does not yield reliable estimates, particularly in the localized phase where the error bars are large. This is due to the fact that in the localized phase Bloch oscillations occur over a limited spatial extension. Therefore, different values of $h$ may localize the excitation at very similar locations at a given time, leading to indistinguishable population distributions and poor sensing. On the other hand, the estimation becomes significantly improved, by incorporating the information obtained from those three times, keeping the total number of measurements $\mathcal{M}$ fixed, as shown in Fig.~\ref{Fig3_BE_1excitation}\textbf{e}. In the SM (see Note 4), we present results for incorporating measurement data from two distinct times. Comparing the results reveals that increasing sampling instances with respect to time can systematically enhance the estimation precision.\\

\begin{figure}[t]
	\centering
	\includegraphics[width=0.9\linewidth]{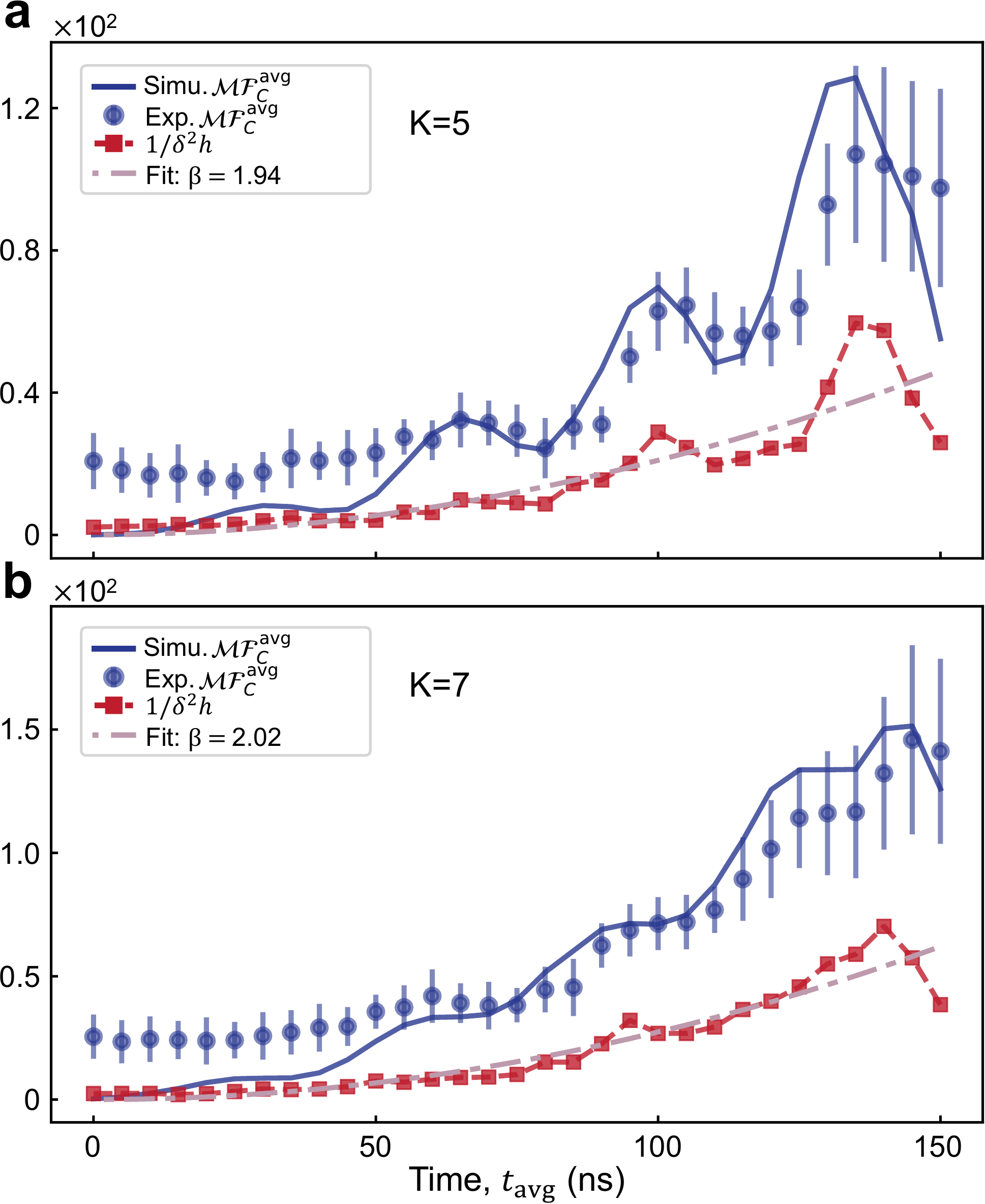}\\
\caption{\textbf{Demonstration of quantum-enhanced sensitivity.} To demonstrate quantum-enhanced sensitivity, the scaling of both the normalized multiple-time averaged CFI, denoted as \(\mathcal{M} \mathcal{F}^{\rm avg}_C\) (expressed in units of $\textrm{MHz}^{-2}$), and the reciprocal of the variance, denoted as \(1/\delta^2 h\), derived from the multiple-time averaged posterior (repeated and averaged over 200 groups), are analyzed as functions of the averaged time. Two different cases are considered: \textbf{a}, \(K {=} 5\) points for time-averaging, with 100 measurement samples taken at each time point, resulting in a total measurement sample size of \(\mathcal{M} {=} 500\). \textbf{b}, \(K {=} 7\) points for time-averaging, with 100 measurement samples taken at each time point, resulting in a total measurement sample size of \(\mathcal{M} {=} 700\). As shown in the figure, \(1/\delta^2 h\) is upper bounded by \(\mathcal{M} \mathcal{F}^{\rm avg}_C\), consistent with the Cramér-Rao inequality. Additionally, the fitting function for \(1/\delta^2 h\) exhibits a \(t^2\) scaling, indicating quantum-enhanced sensitivity achieved through multiple-time point measurement samples. This demonstrates that the multiple-time measurement sample approach can achieve quantum-enhanced sensitivity even when using a simple measurement in the computational basis.}
    \label{Fig4_enhanced}
\end{figure}


\noindent
\textbf{Demonstration of quantum-enhanced sensitivity} \\
\noindent
A significant question is whether our experiment can demonstrate quantum-enhanced sensitivity. Since the system size is fixed in our experiment, quantum-enhanced sensitivity can be considered with respect to time. Indeed, we show that the precision of estimating the parameter $h$ using multiple time points increases super-linearly with time.  Combining the measurement data from $K$ different time points results in a posterior distribution of the form $f(h|\mathbf{n^{(t_1)}},\cdots,\mathbf{n^{(t_K)}}){\sim}\prod_{i{=}1}^K f(h|\mathbf{n^{(t_i)}})$, where $\mathbf{n^{(t_i)}}{=}\{n_j^{(t_i)}\}$ represents the measurement outcomes at time $t{=}t_i$. The additivity of Fisher information implies that $\mathcal{F}_C^{\mathrm{tot}}{=}\sum_{i{=}1}^K \mathcal{F}_C^{(t_i)}$. Therefore, by dividing the total measurement samples $\mathcal{M}$ equally among the different times one can obtain $\delta^2 h{\ge} 1/\mathcal{M}\mathcal{F}_C^{\mathrm{avg}}$, with $\mathcal{F}_C^{\mathrm{avg}}{=}\mathcal{F}_C^{\mathrm{tot}}/K$ being the average of the CFIs. 

We consider $K{=}5$ and $K{=}7$ time points to see a smooth behavior which can be reasonably fitted. In Fig.~\ref{Fig4_enhanced}\textbf{a-b} we plot the rescaled average CFI, namely  $\mathcal{M}\mathcal{F}_C^{\mathrm{avg}}$, obtained from both theory and experiment, as well as  $1/\delta^2h $, obtained from experiment, as a function of $t_{\mathrm{avg}}{=}\frac{1}{K}\sum_{i=1}^K t_i$, for the same linear field gradient $h{=}{-}30$ MHz, with $K{=}5$ and $K{=}7$, respectively. The time points are considered equally spaced with $5$ ns difference between two consecutive points, i.e., $t_{i+1}{-}t_{i}{=}5$ ns, which makes $t_{\mathrm{avg}}{=}t_3$ for $K{=}5$ and $t_{\mathrm{avg}}{=}t_4 $ for $K{=}7$.
As the figure shows, the rescaled average CFI obtained from experiment and theory match well. In addition, as expected, the rescaled average CFI always remains as an upper bound for $1/\delta^2h$. Most importantly, the fitting for $1/\delta^2h{\sim}t^\beta$ shows super-linear behavior  with $\beta{=}1.94$ and $\beta{=}2.02$ for the two values of $K{=}5,7$, respectively. This clearly shows quantum-enhanced precision with simple computational basis measurements. Note that at long evolution times, the decoherence effect becomes important and super-linear scaling deteriorates. This is discussed in the SM, see Note 3, in which we show that as time grows the exponent decreases and eventually reaches sub-linear scaling. \\

\begin{figure}[t]
	\centering
	\includegraphics[width=1\linewidth]{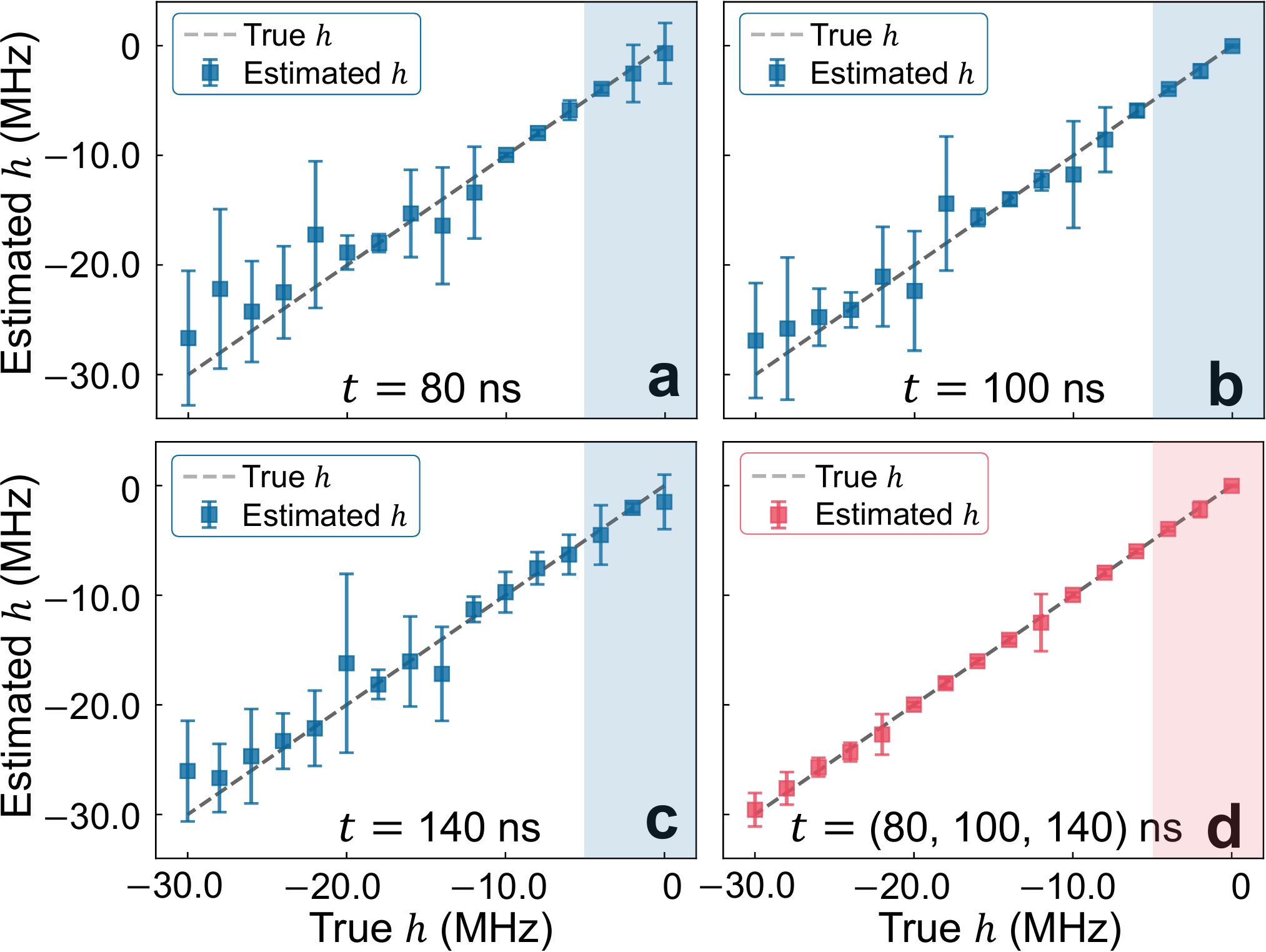}\\
	\caption{\textbf{Double-excitation probe.} Bayesian estimation of parameter $h$ with double excitations in the initial state, using single- and multiple-time averaged approaches, respectively. The estimation is performed for fields ranging from $0$ MHz to ${-}30$ MHz.
    \textbf{a-c}, Single-time estimation using $\mathcal{M}{=}75$ measurement samples at \textbf{a}, $t {=} 80$, \textbf{b}, $t {=} 100$, and \textbf{c}, $t {=} 140$ ns, respectively. \textbf{d}, Multiple-time estimation using $25$ measurement samples at each time  ($t{=}80,100,140$ ns), totaling $\mathcal{M}{=}75$ measurements, matching the single-time cases. In all the panels, the dashed line represents the perfect estimation and the error bars indicate the 1 SD of the estimates obtained from $30$ independent data groups.}
    \label{Fig5_BE_2excitations}
\end{figure}

\noindent
\textbf{Double excitations} \\
\noindent
Furthermore, we extend our investigation to probes with double excitations. Indeed, increasing the number of excitations within the system enhances the probe's precision, as evidenced in our pervious work~\cite{manshouri2024quantum}. The system is initialized in the state $\ket{\Psi(0)} {=} \ket{\mathbf{3,7}}$, where all qubits are prepared in $\ket{0}$ except qubits 3 and 7 in state $\ket{1}$. Subsequently, the system evolves under the Hamiltonian specified in Eq.~(\ref{eq_Ham}) and is measured in the computational basis at time $t$. The measurement is described by projectors $\{ \Pi_{j,k}{=}\ket{\mathbf{j,k}}\bra{\mathbf{j,k}}\}$. In the SM, see Note 4, we present the quantum walk and the Bloch oscillations in the three regimes of the extended phase, around the transition point and the many-body localized phase. The same Bayesian estimation procedure can also be used here. In Fig.~\ref{Fig5_BE_2excitations}\textbf{a-c}, we plot the estimated $h_{\rm est}$ as a function of true value of $h$ for $\mathcal{M}{=}75$ at time $t{=}80$ ns, $t{=}100$ ns and $t{=}140$ ns, respectively. The results demonstrate that, in the localized phase the estimation accuracy is limited and error bars (estimated by 30 repetitions)  are large. On the other hand, by dividing the measurements between the three time points, each using $25$ measurement data while keeping the total measurement $\mathcal{M}{=}75$, one can get a much improved precision shown in  Fig.~\ref{Fig5_BE_2excitations}\textbf{d}. It is evident that aggregating the results from all three time points results in a substantially more accurate estimation, accompanied by markedly reduced error bars.\\

\noindent
\textbf{\large{DISCUSSION}}\\
Our work is a proof of principle experiment which establishes Stark-Wannier localization as a powerful mechanism for quantum-enhanced sensing, leveraging both criticality-enhanced sensitivity and non-equilibrium dynamics to achieve high-precision parameter estimation across a broad range of parameters. Despite the fact that our system size is fixed, the criticality-enhanced sensitivity reveals itself in the magnitude of the error bars which are significantly smaller in the extended phase compared to the localized regime. The advantage of non-equilibrium dynamics is revealed through near-Heisenberg-limited precision with respect to time through combining measurements at different times  using only computational basis measurements, readily implementable across nearly all quantum platforms. This makes Stark-Wannier architecture particularly promising for practical quantum sensing applications across other physical platforms, such as cold atoms and ion-traps, for gravimetry and electric/magnetic field sensing.\\

\noindent
\textbf{\large{METHODS}} \\ 
\textbf{Open quantum system dynamics}\\
\noindent
As a prototypical open quantum system, a qubit experiences energy relaxation and dephasing due to its interactions with the environment, readout resonator, and other control lines, which leads to the decrease in energy relaxation time $T_1$ and dephasing time $T_2^*$ of the qubit. The dynamics of an open quantum system under the Born-Markovian approximation is governed by the Lindblad master equation:
\begin{eqnarray}
   \frac{d\hat{\rho}(t)}{dt} &=& -i\left[\hat{H}, \hat{\rho}(t)\right] + \frac{1}{2T_2^*}\sum_j\left(\hat{\sigma}^z_j \hat{\rho}(t) \hat{\sigma}^z_j - \hat{\rho}(t) \right)  \nonumber  \\  
   &+&  \frac{1}{T_1}\sum_j\left(\po{j}{-} \hat{\rho}(t) \po{j}{+} -\frac{1}{2}\{\po{j}{+} \po{j}{-}, \hat{\rho}(t) \}\right) 
  \label{eq_Lindblad}
\end{eqnarray}
where $\hat{\rho}(t)$ denotes the time-dependent density matrix of the system and $\hat{\sigma}^z_j$ is the Pauli matrix $z$ acting on qubit $j$. The first term of Eq.~(\ref{eq_Lindblad}) describes unitary evolution under the Hamiltonian,  the second term quantifies dephasing and the third term accounts for dissipation.   
The $T_1$ and $T_2^*$ measured at idle points are listed in  Table~\ref{tab1} of the SM. The average $T_1$ ($25.1\,\mu \textrm{s}$) substantially exceeds experimental timescales, whereas the average $T_2^*$ ($1.9\,\mu \textrm{s}$) approaches operational time windows. In practice, interactions at working points may extend the effective dephasing time. By incorporating  dissipation and dephasing for the dynamics of the system, one obtains a good quantitative agreement between experimental estimation of the CFI and its numerical simulation, as shown in Fig.~\ref{Fig2_transport_and_FI}\textbf{g-i}. This shows that dephasing becomes a major limitation factor, in particular, for long time dynamics.  \\ 

\noindent
\textbf{Bayesian analysis} \\
\noindent
The estimation protocol typically involves two steps. First, we need to rebuild the empirical probability distribution $\tilde{P}_j(t,h)$ for all values of $h$ and for a range of time points from experimental sampling data. Each probability at a specific time point is sampled from $4.5\times10^4$ shots for single excitation and $3.5{\times}10^4$ for double excitations. After rebuilding the empirical probability distributions using the gathered data, we collect fresh samples for parameter estimation.
For given measurement outcomes $\{n_j\}$, with the total sample size of $\sum n_j{=}\mathcal{M}$, the likelihood function is then given by
\begin{equation}
    \label{eq:likelihood_methods}
    f(\{n_j\}|h)=\mathcal{M}!\prod_j \frac{\left[\tilde{P}_{j}(t,h)\right]^{n_j}}{n_j!}.
\end{equation}
The likelihood function can be used for obtaining the posterior $f(h|\{n_j\})$ and thus the Bayesian estimation of $h$ as discussed in the main text.

\bibliography{References.bib} 
~\\
\noindent

\noindent
\textbf{\large{ACKNOWLEDGEMENTS}} \\
We acknowledge Zheng-Hang Sun and Xing-Jian He for insightful theoretical discussions and thank Yu-Ran Zhang and Yong-Yi Wang for their helpful discussions on data analysis. The superconducting quantum device was made at the Nanofabrication Facilities at Institute of Physics, Chinese Academy of Science in Beijing. This work was supported by National Natural Science Foundation of China (Grants No.12404578, 12204528, 12247168, 12447184, 92265207, 92365301, T2121001, T2322030, 12050410253, 92065115 and 12274059), the Innovation Program for Quantum Science and Technology (Grant No. 2021ZD0301800), Beijing Nova Program (No. 20220484121, 20240484652) and Beijing National Laboratory for Condensed Matter Physics (2024BNLCMPKF022).\\

\noindent
\textbf{\large{AUTHOR CONTRIBUTIONS}} \\
H.F., A.B., and K.H. supervised the project; A.B. proposed the theoretical idea while H.F., K.H., H.L., and Y.-H.S. discussed and designed the experiment; Z.X. designed and fabricated the quantum device with the help of B.-J.C., G.-H.L., and X.S.; H.L. and K.H. performed the experiment with the assistance of K.X., Y.-H.S., and Z.W.; H.L., Y.Y., and K.H. analyzed the experimental data with the assistance of A.B. and H.F.; Z.-A.W., J.L., Y.Z. and W.-G.M. contributed to the discussion on experimental parameter optimization and analysis; H.L., Y.Y., and K.H. performed the numerical simulations; A.B. and Y.Y. gave theoretical explanations; Y.-H.S., T.-M.L. and J.Z. contributed to the waveform distortion correction; K.X., Y.-H.S., Z.W., K.Z., Y.-S.X., C.-L.D., Y.L., W.-G.M., C.-P.F., J.-C.S., H.-T.L., S.-Y.Z. and Z.-H.L. helped with the experimental setup; A.B., K.H., H.L., Y.Y., and H.F. co-wrote the manuscript, and all authors contributed to the discussions of the results and development of the manuscript.\\

\noindent
\textbf{\large{COMPETING INTERESTS}} \\	
The authors declare no competing interests.

\clearpage 
\onecolumngrid

\begin{center}
    \textbf{\large{\textit{Supplementary Materials for:} \\ \smallskip Non-Equilibrium Criticality-Enhanced Quantum Sensing with Superconducting Qubits}} \\\smallskip
\end{center}

\vspace{0.6cm}
\beginsupplement
\makeatletter
\def\@hangfrom@section#1#2#3{\@hangfrom{#1#2#3}}
\makeatother

\setcounter{section}{0}
\renewcommand{\thesection}{Supplementary Note \arabic{section}}
\renewcommand{\figurename}{Fig.}
\renewcommand{\tablename}{Table}
\renewcommand\thefigure{S\arabic{figure}}
\renewcommand\theequation{S\arabic{equation}}
\renewcommand\thetable{S\arabic{table}}
\tableofcontents

\newpage
\section{Experimental setup}

\subsection{Device Fabrication}

The superconducting quantum device is fabricated using advanced nanofabrication techniques---including electron beam lithography and photolithography---adapted from semiconductor manufacturing. Its scalable architecture, based on Josephson junctions coupled with on-chip capacitors and inductors, enables the coherent quantum operations essential for information processing. This design offers exceptional controllability and extensibility, establishing it as one of the most promising platforms for practical quantum technology applications, such as quantum sensing.

We fabricated a programmable superconducting quantum device comprising 20 transmon qubits arranged in a one-dimensional chain. Each nearest-neighbor qubit pair is capacitively coupled through a flux-tunable coupler, enabling dynamic control of the effective exchange interaction strength $J$ via frequency modulation. Each qubit is coupled to a dedicated $\lambda/4$ readout resonator for dispersive readout, with an individual control line delivering combined XY (high-frequency microwave) and Z (low-frequency flux-bias) signals to manipulate its quantum state and transition frequency, respectively. The tunable couplers mediate programmable qubit-qubit interactions through capacitive coupling, with $J$ adjustable from ${-}20$ MHz to ${\sim}3.5$ MHz via external flux modulation, though they lack direct readout capabilities.

The device is patterned on a single-crystal sapphire substrate (430~$\mu$m thick, $15 {\times} 15$~mm$^2$) and engineered through a multi-step process encompassing chip design, photolithography, thin-film deposition, and reactive ion etching, packaging, followed by cryogenic characterization. Key steps include:  
1. Metallization: sputter deposition of a 100-nm aluminum layer.  
2. Microwave circuitry fabrication: optical lithography (SPR955 resist, 0.70~$\mu$m) and wet etching define coplanar waveguides, control lines, and qubit/coupler capacitors. 
3. Josephson junctions fabrication: electron-beam lithography (MMA/PMMA bilayer resist) followed by double-angle evaporation (65${-}$nm Al at \({+}60^\circ\), oxidation, 100${-}$nm Al at \(0^\circ\)).  
4. Crosstalk suppression: airbridges interconnect ground planes to mitigate parasitic modes.  

\subsection{Cryogenic configuration}
\begin{figure}[b]
	\centering
	\includegraphics[width=0.95\linewidth]{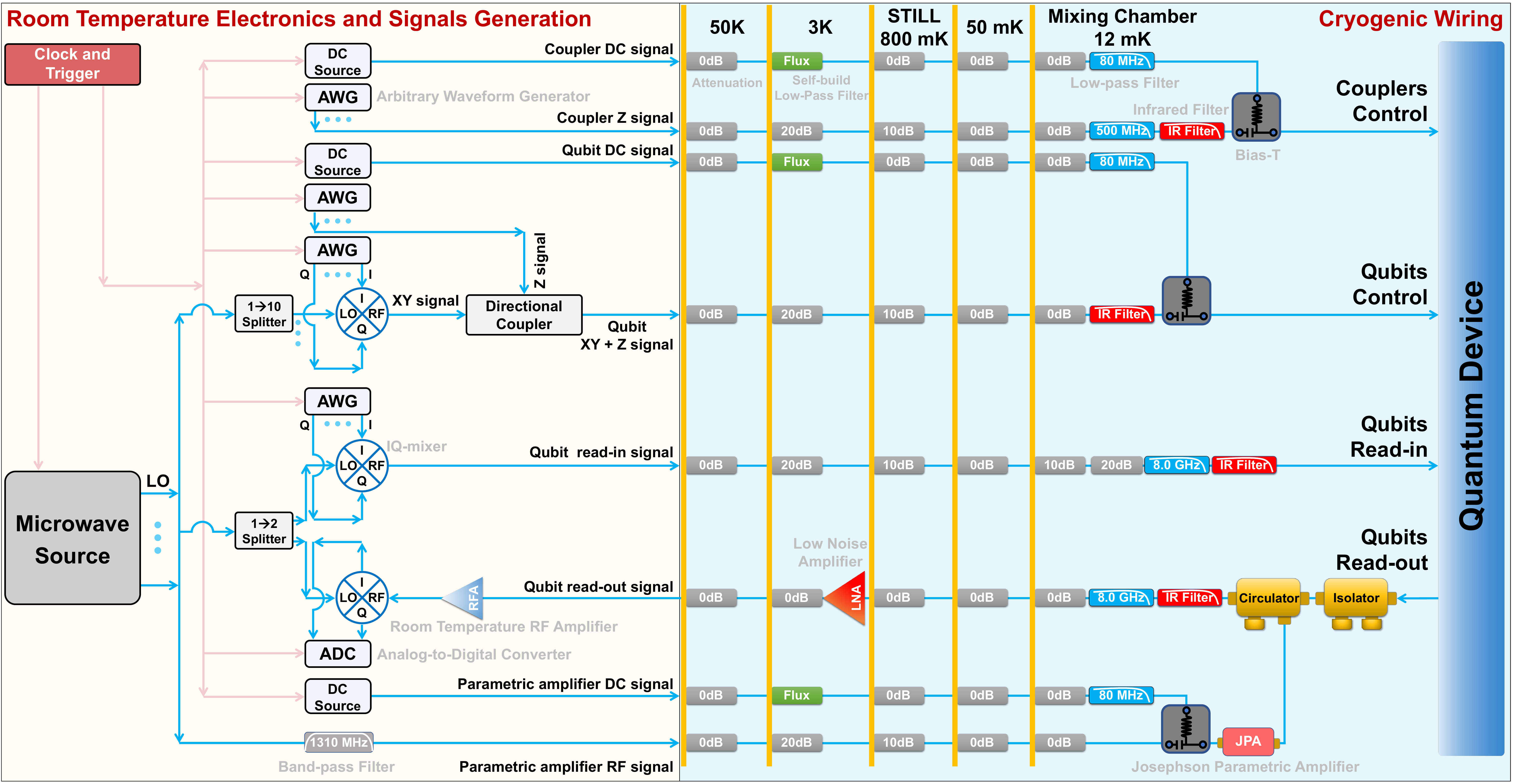}\\
	\caption{\textbf{Room temperature electronics and cryogenic wiring.} The cryogenic wiring connects the quantum device to room-temperature electronics via coaxial transmission lines within a dilution refrigerator, enabling bidirectional signal transmission and maintaining a low-noise environment below 14 mK. The room-temperature electronics, including microwave source, AWG, ADC, etc., are used for high-precision control and measurement of superconducting qubits.}
    \label{FigS1_wiring}
\end{figure}

Achieving high-precision quantum operations requires not only the core superconducting quantum device but also a cryogenic environment, room-temperature electronics for measurement and control, and supporting software systems. The device is fabricated using aluminum and aluminum oxide. Given aluminum's superconducting transition temperature of approximately $1.2\,\mathrm{K}$, the cryogenic environment must maintain temperatures below this threshold. Our transmon-type superconducting qubits operate in the $4{-}7\,\mathrm{GHz}$ frequency range, corresponding to a thermal energy scale of approximately $300\,\mathrm{mK}$. To minimize thermal noise effects, a dilution refrigerator (DR) is employed to reach significantly lower temperatures. Through meticulous cryogenic wiring within the DR, lines serving various functions can be connected to the quantum device, while room-temperature electronics interface with the cryogenic section via coaxial transmission lines, facilitating signal transmission between the cryogenic and room-temperature domains. This architecture enables bidirectional signal transmission between temperature stages. After passing through such extensive control lines, achieving high-precision manipulation of qubits/couplers requires not only optimization of the transmission lines but also proper signal corrections and analyses. These signal processing operations are typically programmed and realized through the client-side software interface. Fig.~\ref{FigS1_wiring} shows the complete configuration of room-temperature electronics and cryogenic wiring used in our experiment. 

Our experimental are performed in a DR (XLD1000) with a base temperature below $7\,\mathrm{mK}$. However, due to the internal wiring arrangement and the introduction of DC signals, the operational environment of the quantum device stabilizes at approximately 14 mK, where thermal excitation effects on the qubits/couplers are almost negligible. The superconducting quantum device is housed in a sample holder enclosed by a custom magnetic shield to suppress electromagnetic interference. XY and Z control signals are coupled at room temperature using directional couplers. Within the DR, from the 50K plate to the sample stage, attenuators (0, 20, 10, 0, and 0 dB) are integrated sequentially to suppress thermal noise (Johnson-Nyquist noise) from each temperature stage. Each qubit and coupler has a dedicated DC control line, with signals passing through custom low-pass flux bias boards and 80 MHz low-pass filters before combining with Z control signals via bias-T at cryogenic temperatures. Readout signals from individual qubit resonators traverse a cryogenic signal chain---including filters, isolators, circulators, Josephson parametric amplifiers, cryogenic low-noise amplifiers, and attenuators---to reach room-temperature electronics with encoded qubit information.

\subsection{Room temperature configuration}
The Z control signals are generated by an arbitrary waveform generator (AWG, MF-AWG-08) with $2\,\mathrm{GS/s}$ sampling rate and $16$-bit vertical resolution. Each Z control output channel includes a DC block ($\sim2\,\mu\mathrm{F}$ capacitor) to suppress low-frequency noise. DC signals originate from a multichannel low-noise DC source (ChipQ-DC-8) with $20$-bit resolution and stability better than $5\,\mathrm{ppm}/10\,\mathrm{h}$. XY control signals and microwave read-in signals are synthesized at room temperature by combining high-frequency carrier signals generated from a multi-channel coherent microwave source (SLFS0218F) with two orthogonal microwave signals from the AWG via an IQ mixer, i.e., microwave orthogonal up-conversion mixing. In practice, the AWG’s DC components and the imperfections of IQ mixer can introduce signal offsets and frequency leakages. These are effectively corrected by adjusting the phase and amplitude of the AWG’s sideband signals and analyzing the output with a spectrum analyzer. Read-out signals carrying qubit state information from the quantum device are acquired and demodulated using an analog-to-digital converter (ADC, MF-DAQ-04) with a sampling rate of 1 Gs/s and 14-bit vertical resolution. A high-stability rubidium clock (STM-Rb-NPA) provides a 10 MHz reference (for the multi-channel coherent microwave source), which is converted to a 250 MHz signal via a clock converter to synchronize the AWG and ADC. Our modular quantum measurement and control software is deployed on a local classical computer, interfacing with a multitude of room-temperature electronic instruments through a high-speed local optical communication network ($10\,\mathrm{Gbps}$).

\subsection{Qubit and coupler charateristics}

We conduct the experiment on a programmable superconducting quantum device consisting of a linear array of 20 transmon qubits interconnected by 19 frequency-tunable couplers. In this work, we employ a subset of 9 qubits (Q$_1$–Q$_9$) and 8 couplers (C$_1$–C$_8$), while the remaining elements are detuned below $3.5\,\mathrm{GHz}$ to suppress unwanted interactions.

To suppress crosstalk, we configure the idle frequencies $\omega_j^{\textrm{idle}}/2\pi$ ($j{=}1,\dots,9$) with nearest- and next-nearest-neighbor detunings exceeding $400\,\mathrm{MHz}$ and $30\,\mathrm{MHz}$, respectively. As shown in Table~\ref{tab1}, the qubits exhibit energy relaxation times $T_{1,j}$ and dephasing times $T_{2,j}^*$ that substantially exceed the system's maximum evolution duration ($350\,\mathrm{ns}$), which mitigates significant decoherence effects throughout the experiment. The direct capacitive coupling strength between each coupler and its adjacent qubits ($g_{j,j}^{Q,C}/2\pi$ and $g_{j+1,j}^{Q,C}/2\pi$) is approximately $90\,\mathrm{MHz}$. To mitigate measurement errors, we implement dynamic readout correction on the observed probability distribution $(P_{0,j},P_{1,j})^\mathrm{T}$ by tracking temporal variations in the readout fidelities of $\ket{0}$ ($F_{0,j}$) and $\ket{1}$ ($F_{1,j}$) states throughout the experiment. This correction applies an inverse transformation derived from the readout fidelity matrix of each addressed qubit Q$_{j}$:
\begin{equation}
\mathcal{M}{_j}^{-1} = \left(\begin{array}{cc}
F_{0,j} & 1-F_{0,j} \\
1-F_{0,j} & F_{1,j} \
\end{array}\right)^{-1}
\end{equation}
where $F_{k,j}$ ($k{=}0,1$) represents the state-dependent measurement fidelity for Q$_{j}$. Additional qubit and coupler parameters are listed in Table~\ref{tab1}.

\begin{table*}[ht]
	\centering
 \caption{\textbf{Qubit and coupler characteristics.} 
$\omega_j^{\textrm{idle}}/2\pi$ is the idle $|0\rangle \rightarrow |1\rangle$ transition frequency of qubit/coupler where decoherence parameters are measured (qubits only): energy relaxation time $T_{1,j}$, Ramsey gaussian dephasing time $T_{2,j}^{*}$, and spin echo dephasing time $T_{2,j}^{\textrm{echo}}$. $T_{2,j}^{*}$ is determined using Ramsey fringe experiments with two $X_{\pi/2}$ pulses, while $T_{2,j}^{\textrm{echo}}$ is characterized by spin echo experiments with one inserted $X_{\pi}$ pulse. $\omega_j^{\textrm{max}}/2\pi$ denotes the maximum frequency and $\eta_{j}/2\pi$ the anharmonicity of qubit/coupler, defined as $\eta_{j} \equiv \omega^{10}_{j} - \omega^{21}_{j}$, where $\omega^{10}_{j}/2\pi$ and $\omega^{21}_{j}/2\pi$ represent the transition frequencies from $|0\rangle$ to $|1\rangle$ and from $|1\rangle$ to $|2\rangle$, respectively. $\omega_j^{\textrm{r}}/2\pi$ is the readout resonator frequency for Q$_{j}$, with $g_{j,r}/2\pi$ the qubit-resonator coupling strength. $F_{0,j}$ and $F_{1,j}$ are the readout fidelities when Q$_{j}$ is prepared in $|0\rangle$ and $|1\rangle$ states, respectively. $g_{j,j}^{Q,C}/2\pi$ and $g_{j+1,j}^{Q,C}/2\pi$ are the coupling strengths between coupler C$_{j}$ and qubits Q$_{j}$, Q$_{j+1}$, obtained from anti-crossing measurements. The effective qubit-qubit coupling $g_{Q_{j},Q_{j+1}}^{\textrm{eff}}/2\pi$ is characterized by the probability exchange-type experiments between Q$_{j}$ and Q$_{j+1}$ with C$_{j}$ at its idle frequency.}
	\label{tab1}
	\begin{ruledtabular}
	\begin{tabular}{ccccccccccc}
		
		Qubit &   Q$_{1}$  &   Q$_{2}$  &  Q$_{3}$  &   Q$_{4}$  &   Q$_{5}$  &   Q$_{6}$  
		 &   Q$_{7}$ & Q$_{8}$ & Q$_{9}$\\
		\hline
		$\omega_j^{\textrm{idle}}/2\pi$ (GHz) & 4.390 & 4.855 & 4.430 & 4.970 & 4.460 & 4.910 & 4.330 & 4.795 & 4.300\\

		$\omega_j^{\textrm{max}}/2\pi$ (GHz) & 4.749 & 5.288 & 4.636 & 5.176  & 4.659 & 5.192 & 4.645 & 5.172   & 4.645\\

            $\omega_j^{\textrm{r}}/2\pi$ (GHz) & 6.631 & 6.657 &  6.667 & 6.698 & 6.720 & 6.735 &  6.751 & 6.781 & 6.803\\

            $\eta_{j}/2\pi$ (GHz) & 0.204 & 0.198 & 0.202 & 0.198  & 0.204 & 0.206 & 0.216 & 0.204   & 0.212\\

            $g_{j,r}/2\pi$ (MHz) & 44.80 & 34.09 & 49.66 & 34.49  & 44.74 & 36.59 & 47.13 & 35.69 & 46.38\\
        
		$T_{1,j}$ ($\mu$s) & 16.1 & 22.9 & 13.7 & 19.4 & 33.1 & 26.8 & 32.9 & 26.8 & 34.2 \\
		$T_{2,j}^*$ ($\mu$s) & 2.1 & 1.6 & 2.0 & 2.1 & 2.2 & 1.5 & 1.9 & 1.5 & 2.1 \\

            $T_{2,j}^{\textrm{echo}}$ ($\mu$s) & 9.6 & 7.9 & 6.0 & 12.3 & 13.8 & 12.1 & 10.1 & 9.8 & 10.2 \\        
		$F_{0,j}$ & 0.952 & 0.972 & 0.938 & 0.956 & 0.967 & 0.987 & 0.959 & 0.968 & 0.948 \\
		$F_{1,j}$ & 0.901 & 0.910 & 0.875 & 0.898 & 0.903 & 0.895 & 0.904 & 0.888 & 0.912\\
		\hline
		\hline
		Coupler &   C$_{1}$  &   C$_{2}$  &  C$_{3}$  &   C$_{4}$  &   C$_{5}$  &   C$_{6}$  
		 &   C$_{7}$ & C$_{8}$\\
		\hline
		$\omega_j^{\textrm{idle}}/2\pi$ (GHz) & 6.263 & 6.206 & 6.073 & 6.199 & 5.995 & 5.925 & 5.822 & 6.163\\

            $\omega_j^{\textrm{max}}/2\pi$ (GHz) & 6.390 & 6.249 & 6.202 & 6.265 & 6.299 & 6.220 & 6.200 & 6.296\\

            $\eta_{j}/2\pi$ (GHz) & 0.283 & 0.309 & 0.278 & 0.308  & 0.322 & 0.302 & 0.280 & 0.313\\

            $g_{j,j}^{Q,C}/2\pi$ (MHz) & 81.0 & 85.5 & 81.5 & 93.5  & 83.0 & 95.2 & 80.2 & 86.2\\

            $g_{j{+}1,j}^{Q,C}/2\pi$ (MHz) & 90.4 & 80.7 & 90.5 & 84.8  & 91.7 & 82.1 & 90.2 & 76.5\\

            $g_{Q_{j},Q{j{+}1}}^\textrm{eff}/2\pi$ (MHz) & 2.98 & 3.37 & 3.37 & 3.40  & 4.05 & 4.62 & 3.78 & 3.66\\
	\end{tabular}
         \end{ruledtabular}
\end{table*}

\section{Calibration procedure}

\subsection{Z crosstalk correction}

\begin{figure*}[t]
	\centering
	\includegraphics[width=0.9\linewidth]{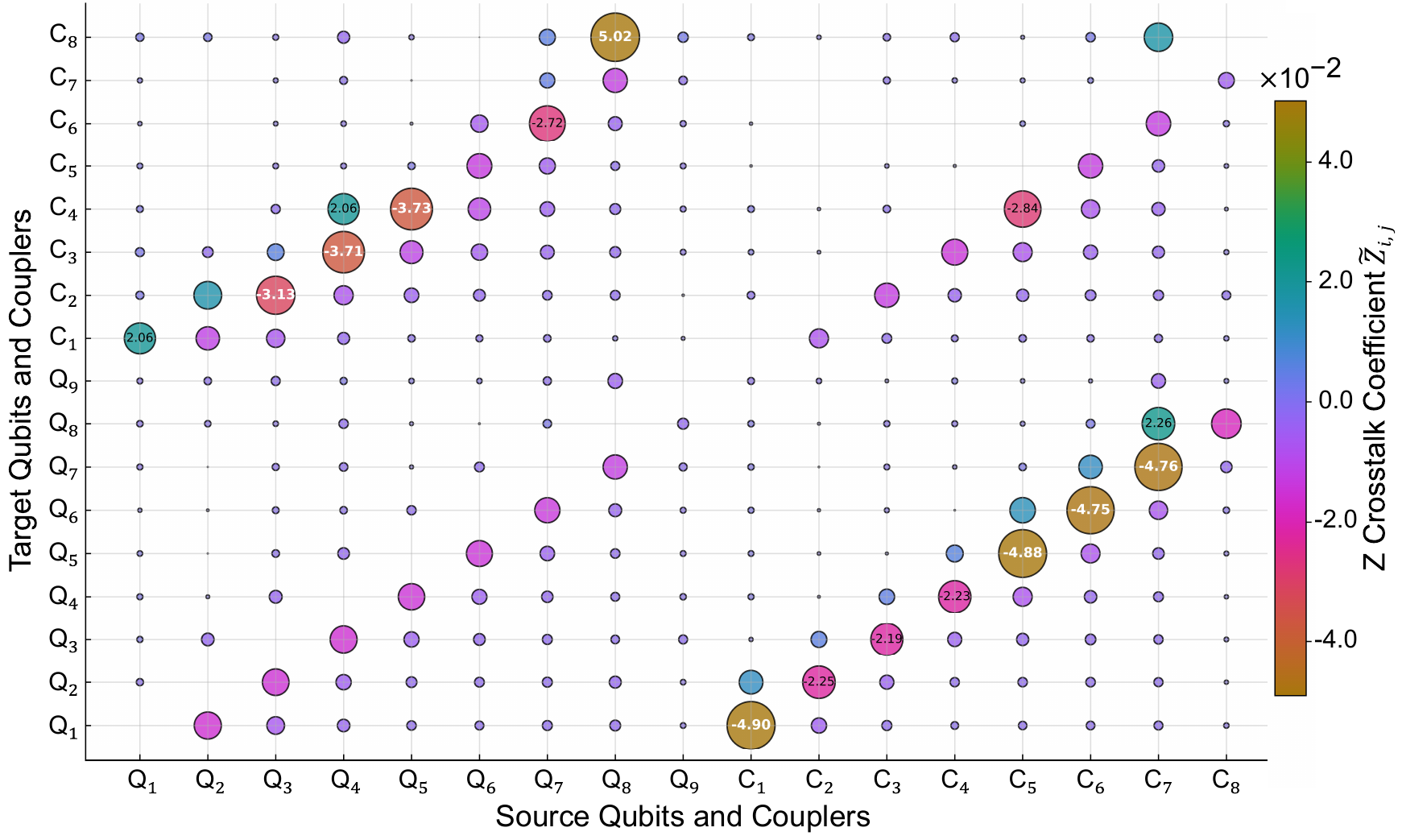}\\
	\caption{\textbf{Off-diagonal elements of the Z crosstalk matrix $\widetilde{Z}$.} Bubble plot visualization of crosstalk coefficients between all experimental qubits and couplers. Each entry $\widetilde{Z}_{i,j}$ quantifies the required Zpa compensation for target qubit/coupler $j$ per unit Z signal applied to source $i$. Bubble sizes scale proportionally to $|\widetilde{Z}_{i,j}|$.}
    \label{FigS2_Zcrosstalks}
\end{figure*}

In this experiment, precise frequency modulation of qubits and couplers via low-frequency Z square-wave signals is essential to achieve uniform coupling strength \(J\) and target gradient field \(h\). However, fabrication imperfections and limited on-chip physical space cause Z control lines of source qubits/couplers to inductively couple to superconducting quantum interference devices (SQUIDs) of unintended targets, generating minor magnetic flux leakage, termed Z crosstalk. When this crosstalk exceeds a threshold (e.g., ${>}$5\%), it significantly compromises multi-qubit control fidelity.

Following the procedures we conducted in Ref.~\cite{li_GAAH_2023}, we experimentally measured Z crosstalk coefficients between all qubits and couplers in the quasi-linear regime, constructing the system-wide crosstalk matrix $\widetilde{Z}$. As shown in Fig.~\ref{FigS2_Zcrosstalks}, the relative large crosstalk coefficients localize predominantly between nearest-neighbor elements, consistent with empirical expectations.

For multi-qubit operations, we pre-calibrate the applied Z pulse amplitude (Zpa) using the relation:
\[
\widetilde{Z}_{\mathrm{corrected}} = \widetilde{Z} \cdot \widetilde{Z}_{\mathrm{applied}},
\]
where \(\widetilde{Z}_{\mathrm{applied}}\) denotes the intended Zpa vector, \(\widetilde{Z}_{\mathrm{corrected}}\) represents the crosstalk-compensated signals, and \(\widetilde{Z}\) is the crosstalk compensation matrix.

\subsection{Z-pulse correction of tunable coupler}
\begin{figure*}[t]
	\centering
	\includegraphics[width=0.9\linewidth]{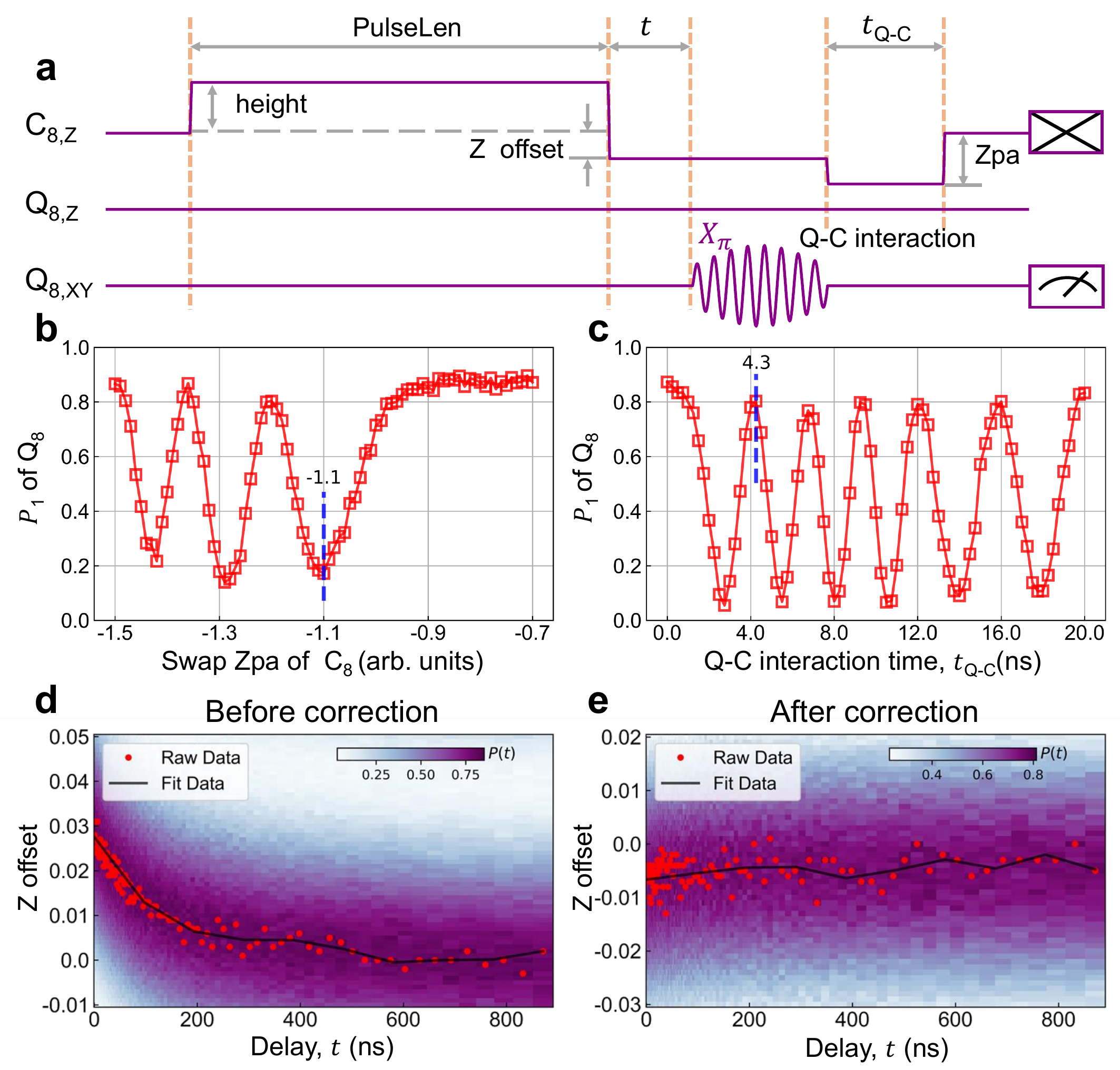}\\
	\caption{\textbf{The calibration of coupler Z pulse.} \textbf{a}, Experimental pulse sequences for calibrating coupler Z pulse shape. \textbf{b}, Resonance $\textrm{Zpa}{=}{-}1.1$ of $\textrm{C}_8$ and \textbf{c}, Resonance time $t_{\textrm{Q}-\textrm{C}}{=}4.3\,\textrm{ns}$ are established at the reference point $(t,\,\textrm{Z}\,\textrm{offset}) {=} (0.0,\,0.0)$ with $\textrm{height}{=}0.0$. \textbf{d}, Waveform distortion of coupler $\textrm{C}_8$ before correction and \textbf{e}, after correction, with the obtained interaction parameters $\textrm{height}{=}{-}1.7$ and $\textrm{PulseLen} {=} 5000\,\textrm{ns}$.}
    \label{FigS3_CouplerPulseCali}
\end{figure*}

Accurate distortion correction of the coupler's Z-control signal is essential for high-precision, stable quantum experiments. However, the absence of an XY-control line and readout resonator precludes direct characterization of coupler signal distortion---the conventional method for qubits~\cite{rol2019fast}. In our previous work, we measured Z-pulse distortion by monitoring qubit excitation population while tuning the qubit near resonance with a coupler, exploiting the frequency shift from their strong transverse coupling~\cite{LiTM2025}. However, this approach requires adjusting qubit energy levels and readout parameters, hindering large-scale, efficient distortion correction.

An alternative experimentally feasible strategy employs neighboring qubit as a probe, where distortion in the coupler Z-control signal is characterized indirectly via state swapping dynamics mediated by the coupler-qubit interaction. Here, we consider $\textrm{C}_8$ and $\textrm{Q}_8$ as an example, with the corresponding waveform sequence shown in Fig.~\ref{FigS3_CouplerPulseCali}\textbf{a}. We first apply a square wave to coupler $\textrm{C}_8$ with sufficiently large height and duration to analyze falling-edge waveform distortion. At the falling edge onset (defined as t = 0), we excite $\textrm{Q}_8$ to |1⟩ by applying an $X_{\pi}$ pulse. The qubit and coupler then undergo state swapping under selected interaction parameters (Zpa, $t_{\textrm{Q}-\textrm{C}}$) chosen to maximize the $|1\rangle$ probability $P_1$ of $\textrm{Q}_8$. However, time-dependent distortion in the falling edge prevents these parameters from maximizing $P_1$ at certain times. To compensate, we apply amplitude corrections (Z offset) at distortion points. By scanning the compensation amplitude until full state exchange is achieved, we indirectly characterize the square wave's distortion profile.

The qubit-coupler interaction parameters are established at the reference point $(t,\,\textrm{Z}\,\textrm{offset}) {=} (0.0,\,0.0)$, with square-wave distortion deliberately excluded during parameters acquisition, i.e., $\textrm{height}{=}0.0$. As shown in Table \ref{tab1}, with $g/2\pi{=}86.2\,\textrm{MHz}$, we initially set $t_{\textrm{Q}-\textrm{C}}{=}\pi/g{=}5.8\,\textrm{ns}$ (Fig.~\ref{FigS3_CouplerPulseCali}\textbf{b}). Scanning $\textrm{Zpa}$ at $t_{\textrm{Q}-\textrm{C}}/2$, we identified the first minimum of $|1\rangle$ state population of $\textrm{Q}_8$ with a target $\textrm{Zpa}{=}{-}1.1$. With $\textrm{Zpa}$ fixed at the target value, we then scan $t_{\textrm{Q}-\textrm{C}}$ and locate a local maximum of $|1\rangle$ state probability of $\textrm{Q}_8$. Among multiple maxima observed in Fig.~\ref{FigS3_CouplerPulseCali}\textbf{c}, we selected $t_{\textrm{Q}-\textrm{C}}{=}4.3\,\textrm{ns}$, which satisfies both minimal duration and maximal $|1\rangle$ probability for $\textrm{Q}_8$. Fig.~\ref{FigS3_CouplerPulseCali}\textbf{d} and \textbf{e} illustrate the waveform distortion of the coupler before and after correction, respectively, with $\textrm{height}{=}{-}1.7$ (around the working point of $\textrm{C}_8$)  and $\textrm{PulseLen}{=}5000\,\textrm{ns}$. Prior to correction, the square wave exhibits significant distortion during the falling edge's initial phase. By determining the Z offset that maximizes $|1\rangle$ probability at each time point and implementing dynamic compensation via a polynomial-exponential hybrid model, effective suppression of waveform distortion is achieved. This result demonstrates the capability of our experimental protocol to correct distortion in coupler Z-control signal.

\subsection{Precise calibration procedure of qubit working frequency}
\begin{figure*}[t]
	\centering
	\includegraphics[width=0.9\linewidth]{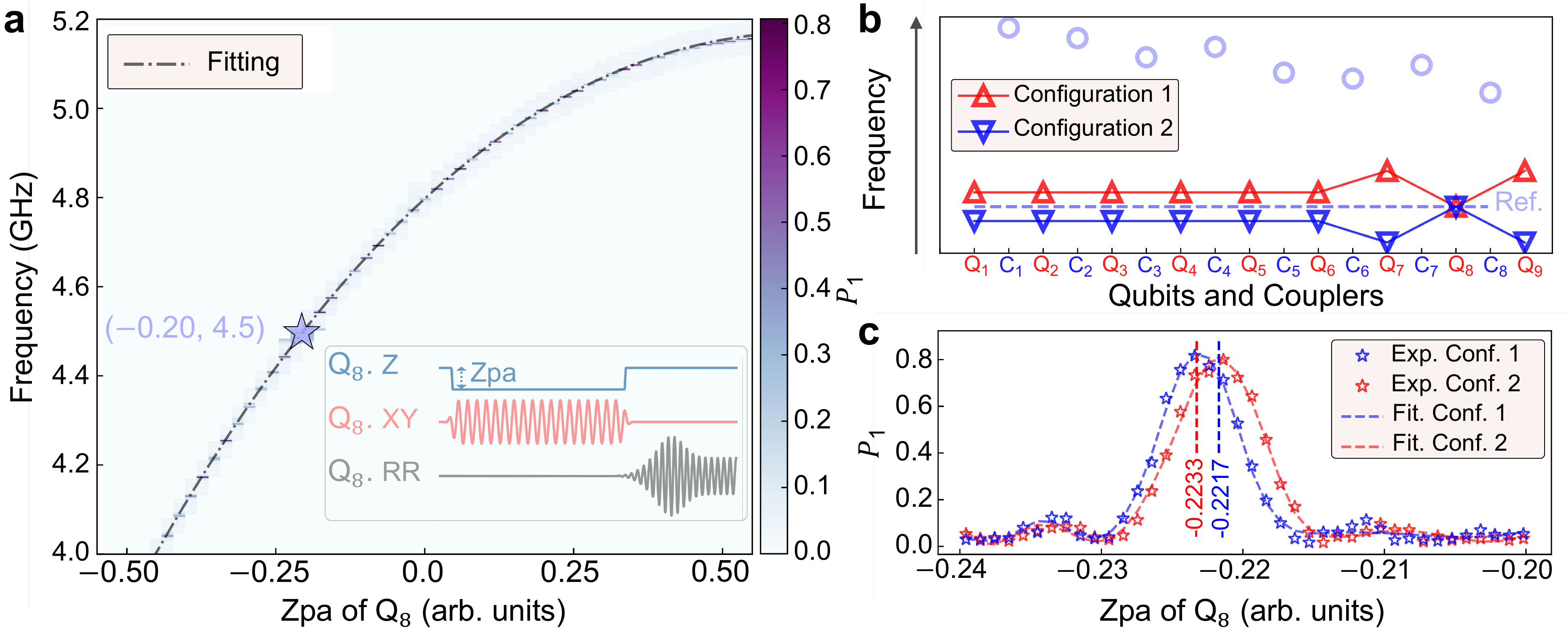}\\
	\caption{\textbf{Precise calibration of qubit working frequency.} \textbf{a}, Experimental automatical energy spectrum of Q$_8$ $|0\rangle {\rightarrow} |1\rangle$ transition, with the inset showing the corresponding waveform sequence. The gray dashed dotted line shows the fit to the experimental data. The light blue pentagon star indicates that 4.5~GHz requires $\textrm{Zpa}{=}{-}0.20$. The XY microwave drive duration used in the experiment is 1.5~\(\mu\)s. \textbf{b}, Two configurations for precisely calibrating Q$_8$ to the reference frequency $ \omega_{\textrm{ref}}/2\pi {=} 4.5~\textrm{GHz}$. \textbf{c}, One-dimensional Rabi oscillation results for both two experimental configurations. The transverse field microwave drive strength is set to approximately 3~MHz, with a drive duration of approximately 80~ns.}
    \label{FigS4_QzpaCali}
\end{figure*}

After completing the Z signal distortion correction for all qubits and couplers, we must precisely calibrate their experimental working points to ensure that we achieve the accurate gradient field \( h \) and uniform coupling strength \( J \) required for our experiments. Here, we use Q$_8$ to demonstrate the qubit working point calibration procedure.

As shown in Fig.~\ref{FigS4_QzpaCali}, with all qubits remaining at their idle points, we scan the  $|0\rangle {\rightarrow} |1\rangle$ transition spectrum of Q$_8$ using the waveform sequence shown in the inset of Fig.~\ref{FigS4_QzpaCali}. Polynomial fitting of the spectrum yields the relationship between Q$_8$'s frequency and applied Zpa—for example, 4.5 GHz requires Zpa ${=} {-}0.20$. After obtaining spectral data for all experimental qubits and couplers, we calibrate working points precisely according to the frequency arrangement in Fig.~\ref{FigS4_QzpaCali}\textbf{b}. To accurately tune Q$_8$ to the reference frequency ($\omega_\textrm{ref}/2\pi{=}$4.5~GHz), we design two frequency configurations—similar to our previous approach in Ref.~\cite{xiang_simulating_2023}—to minimize dispersive and crosstalk effects. In configuration 1, Q$_8$'s nearest neighbors are adjusted to ($\omega_\textrm{ref}/2\pi{+}$100~MHz), while other qubits are adjusted to ($\omega_\textrm{ref}/2\pi{+}$40~MHz). Configuration 2 adjusts these frequencies 100 MHz and 40 MHz below $\omega_\textrm{ref}/2\pi$, respectively. Meanwhile, to minimize coupler influence, we adjust all couplers to their experimental working points based on spectral data. Using this calibration strategy, we fix the drive strength and duration of local transverse ﬁeld,  and calibrate the Zpa corresponding to the highest probability for driving the target qubit to the \( |1\rangle \) state under different configurations (Fig.~\ref{FigS4_QzpaCali}\textbf{c}). The final working parameter is the average of the two optimized Zpa values. Notably, the final optimized Zpa shows approximately 20 MHz frequency shift from the initial spectrum (Fig.~\ref{FigS4_QzpaCali}a), which would significantly impact experimental results if uncorrected.

Moreover, we can select some characteristic frequency points within the experimental working range and repeat the calibration procedure, establishing a parameter mapping that provides the foundation for the subsequent precise calibration of the qubit working parameters.

\subsection{Precise calibration procedure of coupler working frequency}
\begin{figure*}[t]
	\centering
	\includegraphics[width=0.9\linewidth]{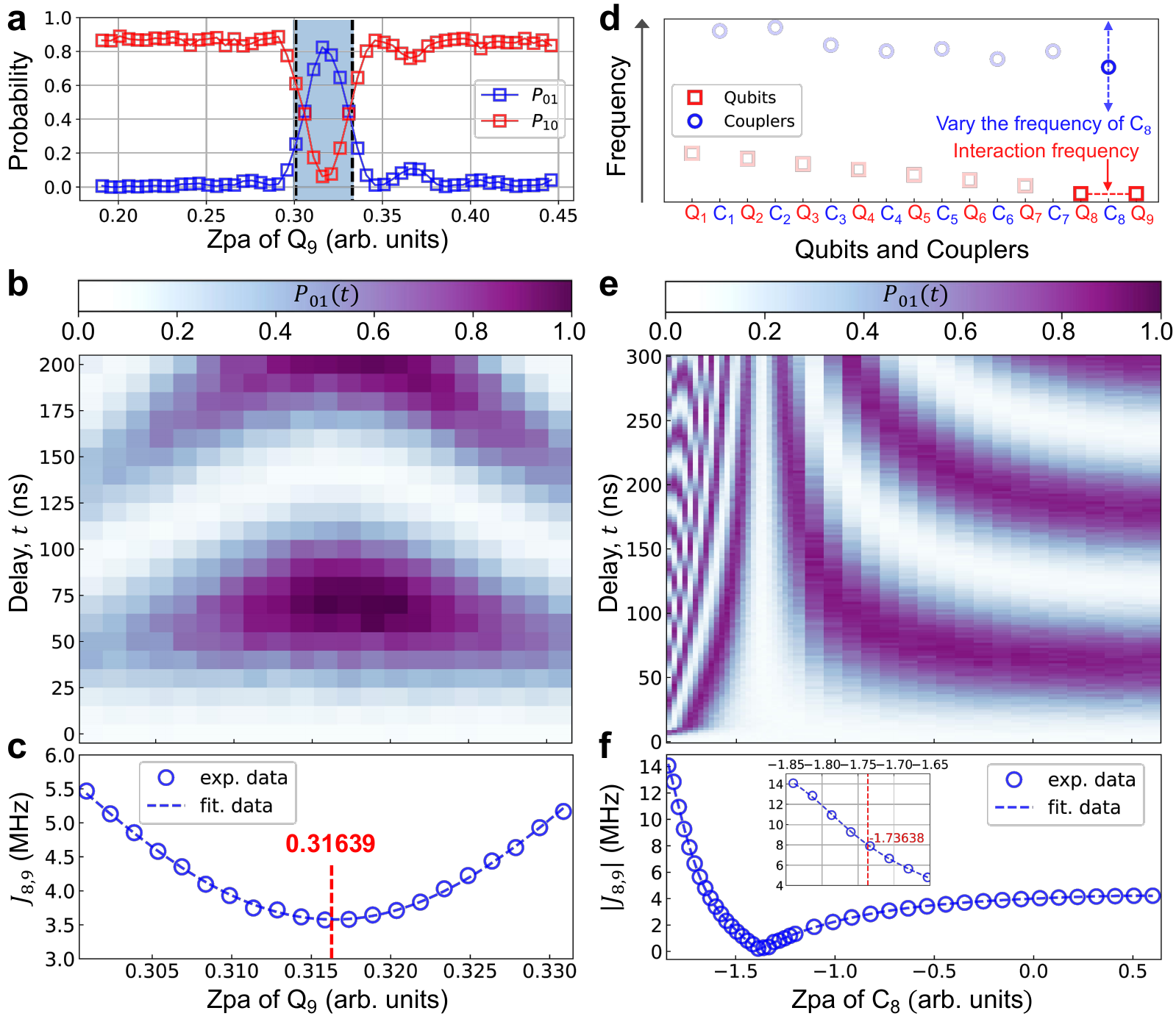}\\
	\caption{\textbf{Precise calibration of coupler working frequency.} \textbf{a},~Resonance parameters calibration of Q$_8$ and Q$_9$. With coupler C$_8$ remaining at idle frequency, the joint probabilities $P_{01}$ (blue squares) and $P_{10}$(red squares) are measured versus Q$_9$'s Zpa with Q$_8$'s Zpa and interaction time (${\sim}60$ ns) fixed. Pronounced population exchange occurs in the light-blue region. \textbf{b}, Detailed scan of the exchange region: time-dependent probability $P_{01}(t)$ versus Q$_9$'s Zpa. \textbf{c}, Fourier transform along the time axis of \textbf{b}, yielding the effective coupling strength $J_{8,9}$ versus Q$_9$'s Zpa. The fit (blue dashed line) to experimental data (blue circles) gives optimal exchange Zpa = 0.31639 for Q$_9$. \textbf{d},~Coupler working point calibration scheme. With optimized Zpas from \textbf{c} applied to Q$_8$ and Q$_9$, the coupler frequency is tuned to achieve target coupling strength. Meanwhile, other qubits and couplers are arranged according to the actual experimental configuration to obtain more accurate parameters corresponding to experiments performed in the main text. \textbf{e}, Time evolution of $P_{01}(t)$ for Q$_8$ and Q$_9$ versus C$_8$ Zpa. As C$_8$ shifts from its idle point toward qubit frequencies, the qubits first experience an exchange blockade region (i.e., decoupling Zpa ${\sim}{-}1.38$ for C$_8$) followed by rapid growth of $|J_{8,9}|$. \textbf{f}, Fourier transform along the time dimension of \textbf{e}, extracting $J_{8,9}$ versus C$_8$'s Zpa. Inset: Fitting the blue circles gives the optimal Zpa (1.73638) of C$_8$ for $|J_{8,9}| {=} 8$ MHz.}
    \label{FigS5_CzpaCali}
\end{figure*}

The experiment relies on the precise frequency control of qubits and couplers to achieve the target Stark-Wannier gradient field $h$ and the nearest-neighbor qubit effective coupling strength $J_{j, j{+}1}$. When other qubits and couplers are remain at their idle points, we can obtain the relationship between the frequency of a target qubit or coupler and its Zpa through simple spectral measurements. However, this mapping does not account when there are residual Z crosstalk and dispersive interactions caused by other qubits and couplers near the experimental working points. Therefore, taking the coupler C$_8$ between Q$_8$ and Q$_9$ as an example, we designed a precise calibration scheme for the coupler's working parameters, as shown in Fig.~\ref{FigS5_CzpaCali}.

As shown in Fig.~\ref{FigS5_CzpaCali}\textbf{a}, based on the relationship obtained from the precise calibration of the qubit working parementers, we fix the Zpa of Q$_8$ to the parameter corresponding to the average working frequency of Q$_8$ and Q$_9$ according to the target Stark-Wannier gradient field $h$. Meanwhile, we maintain the interaction time (${\sim} 60$ ns) and scan the Zpa of Q$_9$ within an empirical range to obtain the probability evolutions $P_{01}$ and $P_{10}$ of the two-qubit states $\ket{01}$ and $\ket{10}$. It is evident that within the Zpa range of 0.3 to 0.33 for Q$_9$ (light blue shaded region), the two states undergo population exchange. We then measure the detailed time-dependent probability evolution $P_{01}(t)$ of the two-qubit state within this region, obtaining the oscillation heatmap shown in Fig.~\ref{FigS5_CzpaCali}\textbf{b}. By performing Fourier transform along its time axis, we obtain the coupling strength data as shown in Fig.~\ref{FigS5_CzpaCali}\textbf{c}. A binomial fit to this data reveals that the Zpa corresponding to the most complete population exchange, i.e., the minimum coupling strength $J_{8,9}$, is 0.31639. It is important to note that during the experiments depicted in Fig.~\ref{FigS5_CzpaCali}\textbf{a} and Fig.~\ref{FigS5_CzpaCali}\textbf{b}, all other non-participating qubits and couplers were approximately set to the working points required for the experiments under the same conditions as in the main text (i.e., with the same Stark-Wannier gradient field $h$).

Subsequently, based on the energy level diagram shown in Fig.~\ref{FigS5_CzpaCali}\textbf{d}, where all other qubits and couplers are near their experimental working points, we perform precise calibration of the coupler's working parameters. We first fix the parameters of Q$_8$ and Q$_9$ determined in the previous steps and then scan the Zpa of C$_8$ within an empirical range to obtain the evolution of $P_{01}(t)$ with C$_8$'s Zpa, as shown in Fig.~\ref{FigS5_CzpaCali}\textbf{e}. From the Fourier transform along the time dimension (Fig.~\ref{FigS5_CzpaCali}\textbf{f}), we observe a clear modulation of the absolute value of effective coupling strength $|J_{8,9}|$. By performing a polynomial fit near the target coupling strength $|J_{8,9}| {=} 8$ MHz, we can obtain the precise working parameter for C$_8$.

In our experiment, a wide range of Stark-Wannier gradient fields $h$ is involved. Therefore, we can select some specific values to repeat the above calibration procedure, thereby obtaining the correspondence between each coupler's experimental parameters and Stark-Wannier gradient field $h$. These steps improve calibration efficiency while ensuring parameter accuracy.

\section{Numerical results}

\subsection{Identifying the transition point}
To determine the field gradient $h$ at which the system goes through a phase transition from the extended to the localized phase, we analyze the time-normalized quantum Fisher information, $\mathcal{F}_Q/t^2$, for several system sizes $L$, as performed in Ref.~\cite{manshouri2024quantum}. In general, the QFI scales as ${\sim} t^2L^\beta$, specifically in the Stark-Wannier model, the exponent $\beta$ depends on the phase: in the extended phase, quantum-enhanced precision can be achieved, namely $\beta {>} 1$, while in the localized phase, it becomes size-independent with $\beta {=} 0$. This distinct scaling behavior manifests as a peak in the normalized QFI when plotted against the gradient field $h$—the normalized $\mathcal{F}_Q/t^2$ reaches its maximum at the phase transition point where the system exhibits critical behavior and enhanced quantum correlations. By dividing the raw QFI by $t^2$, we remove the Heisenberg scaling time dependence, isolating the dependence on $h$ and many-body correlations. We therefore use this peak in $\mathcal{F}_Q/t^2$ as an indicator to identify the critical gradient $h_c$ that separates the extended and localized phases.

\begin{figure*}[t]
	\centering
	\includegraphics[width=0.9\linewidth]{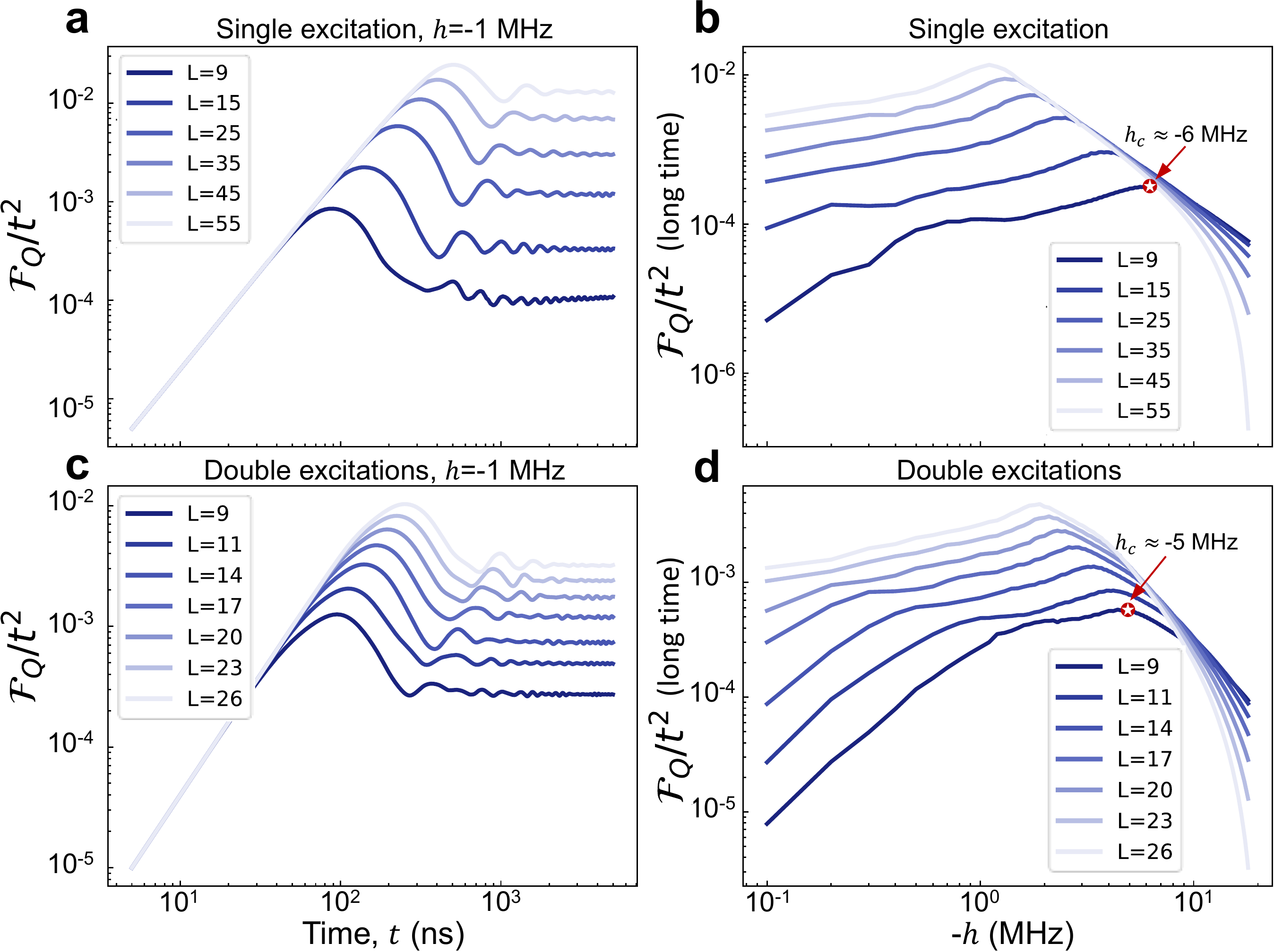}
	\caption{\textbf{Transition point analysis.} \textbf{a}, \textbf{c}, The normalized QFI $\mathcal{F}_Q/t^2$ (in units of $(\textrm{MHz}\cdot\textrm{ns}) ^{-2}$) as a function of time $t$ with different system sizes $L$ for single- and double-excitation scenarios, respectively. As time increases, the normalized quantum Fisher information converges to a fixed value. \textbf{b}, \textbf{d}, Long-time normalized QFI as a function of gradient field strength $h$ for different system sizes in single- and double-excitation cases. The peak of each curve indicates the transition point. For single- and double-excitation initial states with system size $L=9$, the transition point occurs at approximately -6 MHz and -5 MHz, respectively, as marked by the red arrows.}
    \label{figS6_transition_point}
\end{figure*}

 We conducted numerical simulations for various system sizes and gradient field $h$. For system size $L{=}9$, the initial state is set to $|\textbf{5}\rangle$ for single excitation and $|\textbf{3}, \textbf{7}\rangle$ for double excitations. For other system sizes, single excitation is placed at the center of the chain, while the double excitations are positioned at one-third and two-thirds along the chain. In Fig.~\ref{figS6_transition_point} $\textbf{a}$ and $\textbf{c}$, we plot $\mathcal{F}_Q{/}t^2$ as a function of time $t$ for different system sizes, starting from single- and double-excitation initial states, respectively. As the figures show, as time $t$ increases, the normalized QFI gradually converges to a steady value. In Fig.~\ref{figS6_transition_point} $\textbf{b}$ and $\textbf{d}$, we plot the large-time value of $\mathcal{F}_Q{/}t^2$ as a function of ${-}h$ for various odd system sizes, starting from single- and double-excitation initial states, respectively. As $|h|$ increases, the normalized QFI initially rises, reaches a maximum, and then decreases. This peak identifies the critical gradient $h_c$ that marks the phase transition between extended and localized regimes. For single- and double-excitation initial states, using the experimental parameters with $L {=} 9$, we find $h_c {\approx} {-}6$ MHz and $h_c {\approx} {-}5$ MHz, respectively.

\subsection{Decoherence against quantum-enhanced sensitivity}

\begin{figure*}[t]
	\centering
	\includegraphics[width=0.9\linewidth]{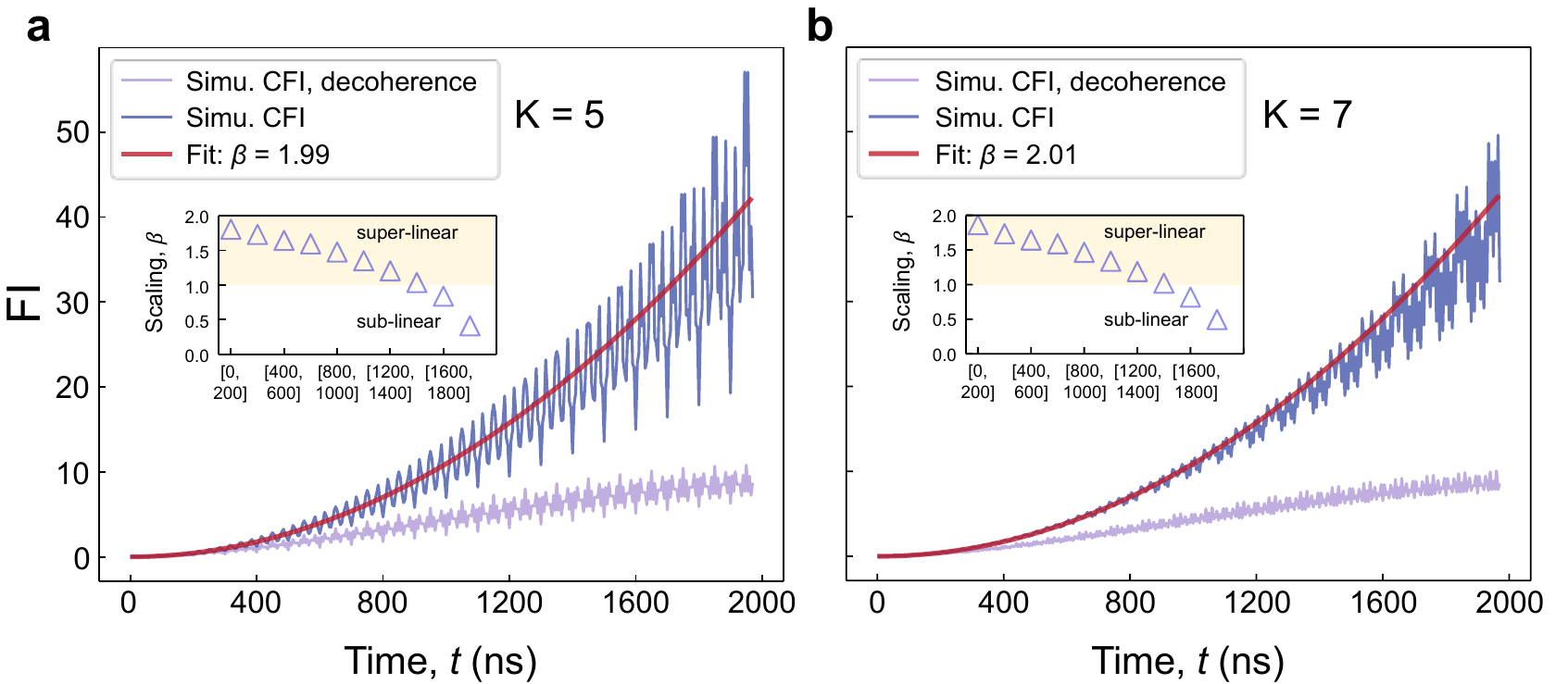}\\
	\caption{\textbf{Simulation of decoherence-induced degradation on scaling.} Fisher information for time-averaging windows \textbf{a}, $K{=}5$ and \textbf{b}, $K{=}7$ at $h{=}{-}30\,\textrm{MHz}$. Blue solid lines represent decoherence-free simulations, achieving power-law exponents $\beta {=} 1.99$ for $K{=}5$ and $2.01$ for $K{=}7$. Purple solid lines denote simulations under experimentally measured decoherence parameters. Insets illustrate the time evolution of the power-law exponent, which exhibits a monotonic decay behavior.}
    \label{FigS7_CFI_decoherence}
\end{figure*}

In the main text, we experimentally demonstrate high-precision quantum sensing that the reciprocal of the variance scales as ${\sim}1/t^\beta$ with $\beta{\simeq} 2$ (or equivalently Fisher information scales super-linearly as $\mathcal{F}^{\rm avg}_C{\sim} t^\beta$) , which exceeds the classical shot-noise scaling ${\sim} 1/t$. However, in open quantum systems, decoherence effects are induced during long evolution and significantly degrade the estimation precision, leading to the loss of quantum-enhanced precision.

In Figs.~\ref{FigS7_CFI_decoherence}\textbf{a-b}, we numerically show how decoherence affects super-linear scaling of CFI $\mathcal{F}^{\rm avg}_C$ under gradient field $h{=}{-}30\,\textrm{MHz}$, with multiple-time averaging over $K{=}5$ and $K{=}7$ time points, respectively. The blue solid lines exhibit the evolution of $\mathcal{F}^{\rm avg}_C$ in the absence of decoherence with a system size of $L{=}9$. The red lines represents the fitting of the multiple-time averaged CFI, yielding a power-law behavior of $\mathcal{F}^{\rm avg}_C {\sim} t^\beta$ with the scaling exponent $\beta{=}1.99$ for $K{=}5$ and $\beta{=}2.01$ for $K{=}7$, respectively. This clearly shows quantum-enhanced sensitivity in a unitary evolution which is matched well with our experimental results within short time scale as shown in the main text. However, by incorporating decoherence effects such as dephasing and energy dissipation (with the same values as discussed in the Methods section of the main text), the evolution of $\mathcal{F}^{\rm avg}_C$ becomes different (purple lines). Although the results show good agreement with unitary evolution at short time scales ($t{<} 200\,\textrm{ns}$), the scaling behavior progressively deteriorates with increasing evolution time. To investigate the temporal scaling dynamics, we analyze multiple $200\,\textrm{ns}$ time windows and extract the scaling exponent $\beta$ through numerical fitting, as shown in the insets of Figs.~\ref{FigS7_CFI_decoherence}\textbf{a-b}. A distinct transition from superlinear ($\beta {>} 1$) to sublinear ($\beta {<} 1$) scaling is observed, indicating the gradual disappearance of quantum-enhanced sensitivity.

\section{Additional experimental data}
\subsection{Two-time combined Bayesian estimation}
\begin{figure*}[t]
	\centering
	\includegraphics[width=1\linewidth]{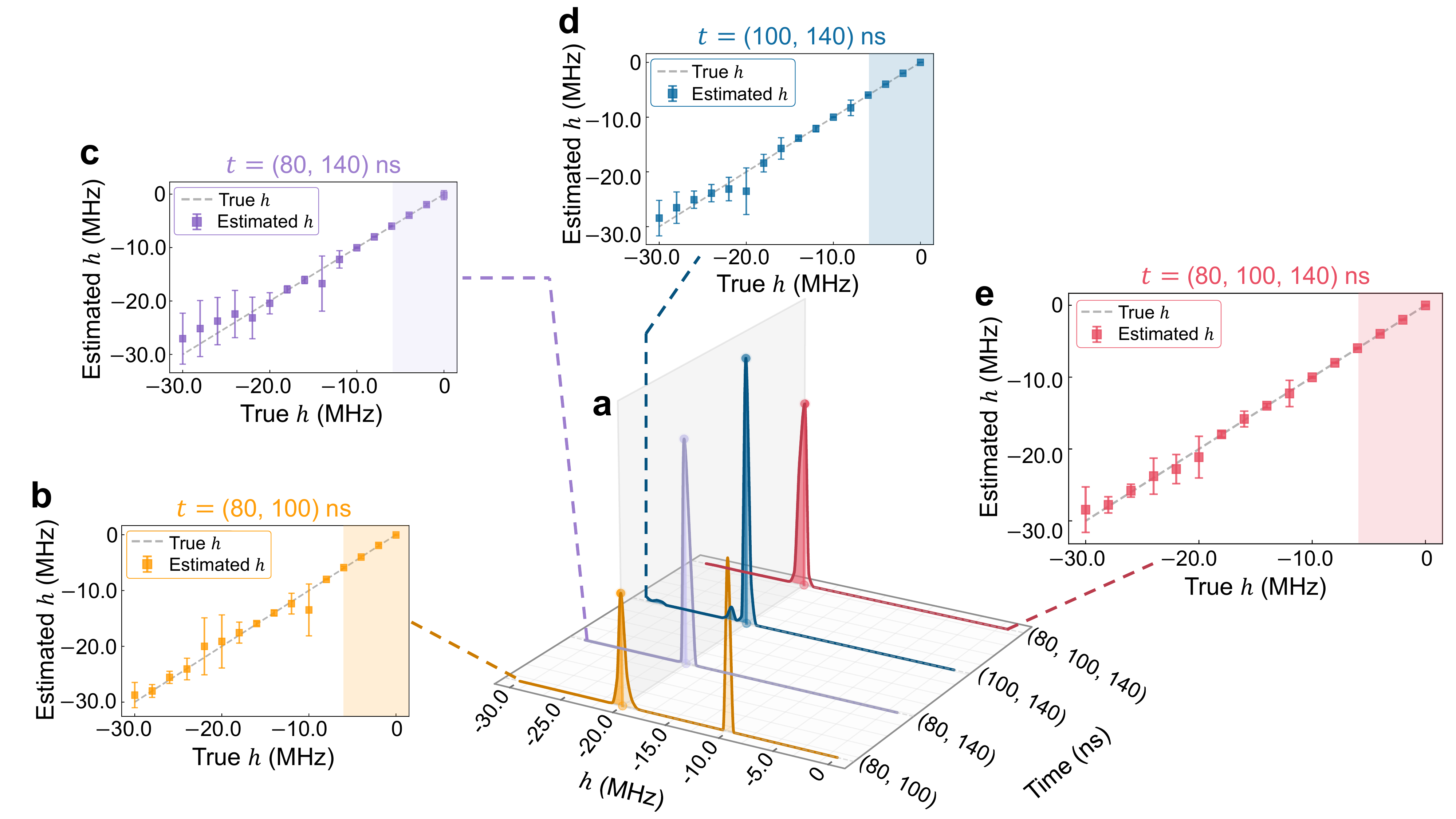}\\
	\caption{\textbf{Bayesian estimation of parameter $h$ using two-time and three-time combined approaches.} \textbf{a}, Two-time posterior probability distributions at  $t\,{=}\,(80,\,100)$ (yellow line), $(80,\,140)$ (purple line), $(100,\,140)\,\textrm{ns}$ (blue line) and three-time posterior probability distribution at $t\,=\,(80,\,100,\,140)\,\textrm{ns}$ (red line). \textbf{b-d}, Correspondence between the estimated value $h_{\rm est}$ and the true $h$, using one-time posterior, within the range of ${-}30\,\textrm{MHz}$ to $0\,\textrm{MHz}$, with error bars representing the 1 SD. The dashed lines represent ideal estimation and the shaded regions denote extended phase. \textbf{e}, The estimated $h$ versus true $h$ for the three-time averaged approach. The total number of measurement samples is $\mathcal{M}{=}60$. For the two-time averaged approach, we partition the measurements samples into two groups ($30$ per group), while for the three-time averaged approach, each group contains $20$ samples.}
    \label{FigS8_BE_1excitation_comparison}
\end{figure*}

As shown in Fig.~\ref{Fig3_BE_1excitation} in the main text, the measurement samples $\mathcal{M}$ may generate a multi-peak structure in the single-time posterior distribution, severely degrading the accuracy of Bayesian estimation. In contrast, the three-time combined approach converges to a sharply unimodal distribution, significantly improving estimation precision and underscoring the pivotal role of multi-time combined approach in mitigating statistical fluctuations. Morover, in Fig.~\ref{FigS8_BE_1excitation_comparison}, we demonstrate the results using two-time combined approach (time pairs: $(80,\,100)\,\textrm{ns}$, $(80,\,140)\,\textrm{ns}$ and $(100,\,140)\,\textrm{ns}$). Although superior to the single-time approach, the results exhibit broader distribution widths and residual multi-peak features compared to the three-time combined approach, resulting in systematic deviations, particularly in the localized regime. The comparative results above clearly demonstrate that incorporating multi-time combined approach contributes to enhancing the accuracy of parameter estimation.

\subsection{Quantum walks of double-excitation probe}
\begin{figure*}[t]
	\centering
	\includegraphics[width=0.9\linewidth]{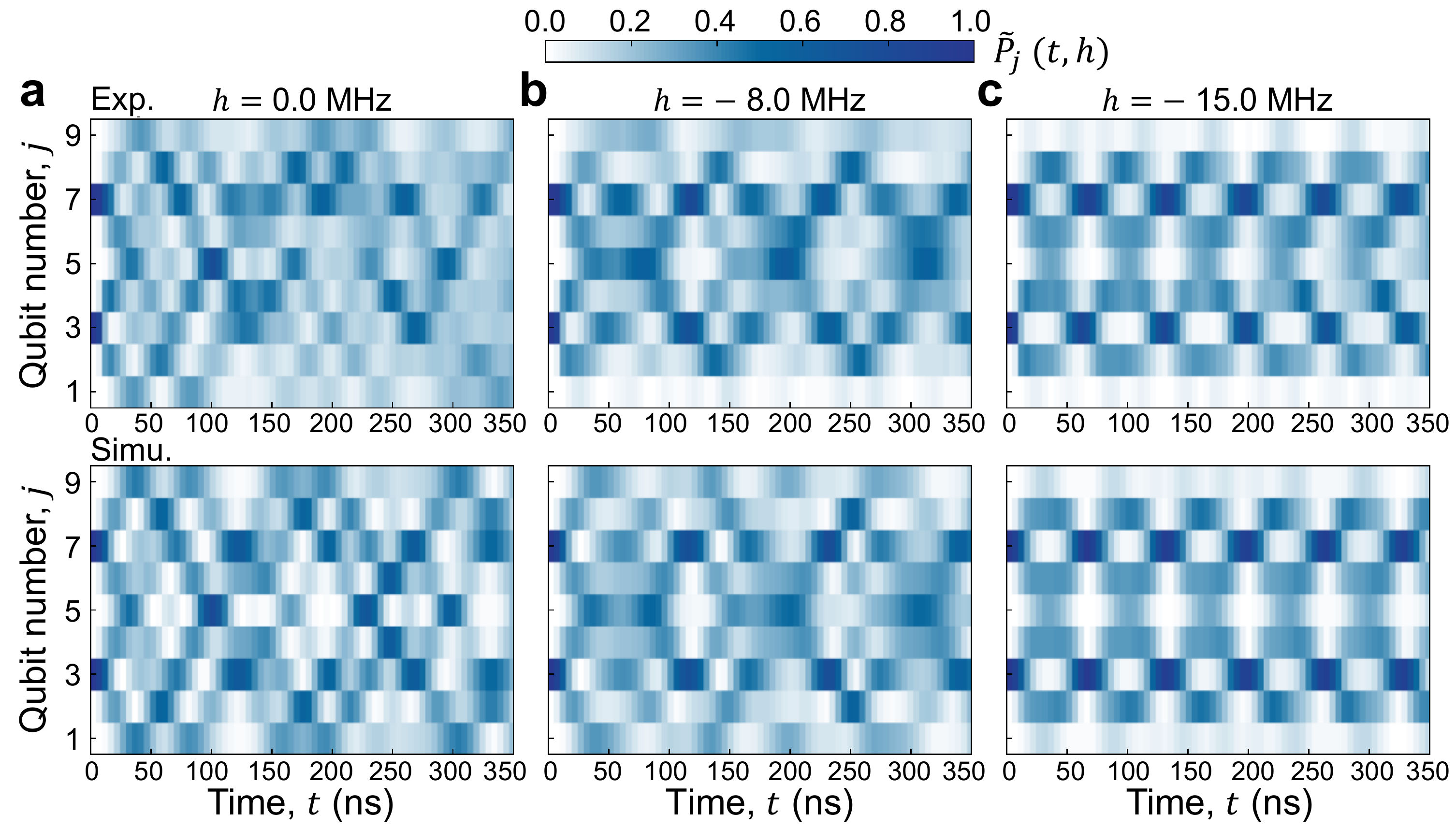}\\
	\caption{\textbf{Particle transport with double excitations.} The time evolution of on-site population $\tilde{P}_j(t, h)$ as the system initialized in $|\Psi(0)\rangle {=} \mathbf{|3, 7}\rangle$ \textbf{a}, \textbf{d}, in extended phase, with $h {=} 0.0\,\textrm{MHz}$; \textbf{b}, \textbf{e}, around the transition point, with $h {=} {-}8.0\,\textrm{MHz}$; \textbf{c}, \textbf{f}, in localized phase, with $h {=} {-}15.0\,\textrm{MHz}$. The upper panel shows averaged experimental data from $10$ independent repetitions with each repeition covers 5000 measurement outcomes, while the lower panel displays numerically simulated results derived from our theoretical framework.}
    \label{FigS9_QW_2excitations}
\end{figure*}

\begin{figure*}[t]
	\centering
	\includegraphics[width=0.9\linewidth]{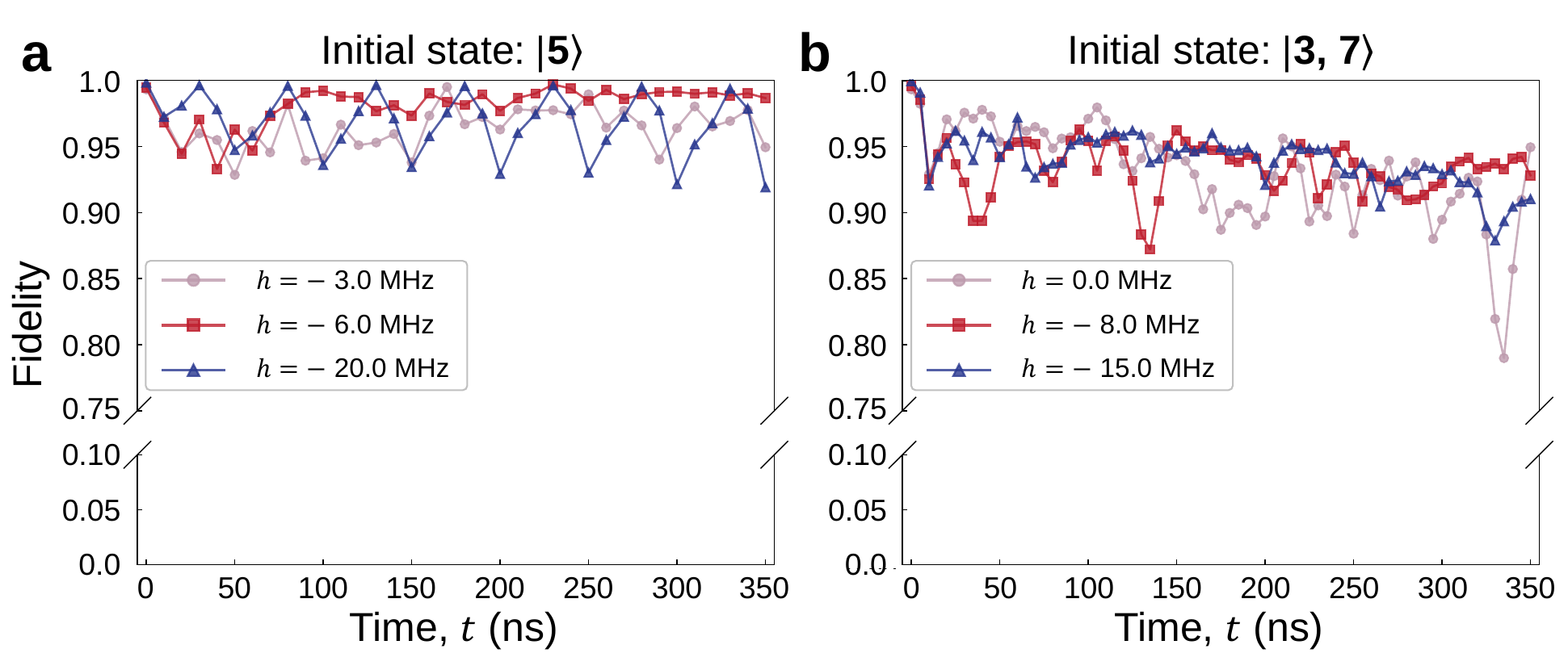}\\
	\caption{\textbf{The performance of experimental quantum walks.} The fidelity of the experimental and theoretical normalized probability distributions of \textbf{a}, single excitation and \textbf{b}, double excitations.}
    \label{FigS10_QW_fidelity}
\end{figure*}

We focus on probes with double excitations with the initial state being $|\textbf{3}, \textbf{7}\rangle$. In Fig.~\ref{FigS9_QW_2excitations}, We present the time evolution of population distribution $\tilde{P}_j(t, h)$ for $h{=}0\,\textrm{MHz}$ (extended phase), $h{=}{-}8\,\textrm{MHz}$ (around the transition point) and $h{=}{-}15\,\textrm{MHz}$ (localized phase) over a $0{-}350\,\textrm{ns}$ evolution window. In the extended phase, the particle rapidly propagates throughout the whole system, with only a small fraction returns to the original site after boundary reflections. Near the critical point, the system's dynamics undergo a dramatic change, marked by the onset of Bloch oscillations. Conversely, in the localized phase, the excitation remains strongly localized around its initial position, exhibiting Bloch oscillations confined to this vicinity without significant diffusion. Upon approaching the critical point, the system exhibits a dynamical transition characterized by the emergence of Bloch oscillations. Within the localized phase, excitations maintain strong spatial confinement near their initial positions, manifesting as Bloch oscillations with negligible diffusion beyond this localized regime.
To quantify the fidelity of the quantum walk evolution, we define the state fidelity at discrete time $t$ as 
\begin{equation}
   F(t) =\sum_j \sqrt{p_j(t)q_j(t)},
  \label{eq_fidelity}
\end{equation}
where $p_j(t)$ and $q_j(t)$ represent the theoretically simulated and experimentally measured probability distributions (normalized) of $\textrm{Q}_j$ in state $|1\rangle$ respectively~\cite{xiang_simulating_2023}. As shown in Fig.~\ref{FigS10_QW_fidelity}, the high fidelity (above $0.92$ and $0.8$ before $t{=}350\,\textrm{ns}$ for single excitation and double excitations respectively) for all linear gradient field $h$ demonstrates well agreement between the experimental results and theoretical simulations.

\end{document}